\documentclass[aps,prc,twocolumn,superscriptaddress,floatfix]{revtex4-1}
\usepackage{textcomp}
\usepackage{nicefrac}
\usepackage{longtable}
\usepackage{graphicx}
\usepackage{multirow}
\usepackage{amsmath}
\usepackage[breaklinks=true]{hyperref}
\usepackage{breakcites}

\begin{document}
 
 \title{Alpha clustering in $^{28}$Si probed through the identification of high-lying $0^+$ states}
 
 \author{P. Adsley}
 \email{padsley@gmail.com}
 \affiliation{Department of Physics,  University of Stellenbosch, Stellenbosch, South Africa}
 \affiliation{iThemba Laboratory for Accelerator Based Sciences, Somerset West 7129, South Africa}
 
 \author{D.G. Jenkins}
 \affiliation{Department of Physics, University of York, Heslington, York, YO10 5DD, United Kingdom}
 
 \author{J. Cseh}
 \affiliation{Institute for Nuclear Research, Hungarian Academy of Sciences, Debrecen, Pf. 51, 4001, Hungary-4001}
 \author{S.S. Dimitriova}
 \affiliation{Institute for Nuclear Research and Nuclear Energy, Bulgarian Academy of Sciences, 1784 Sofia, Bulgaria}
 
 \author{J.W. Br\"{u}mmer}
 \author{K.C.W. Li}
 \affiliation{Department of Physics, University of Stellenbosch, Stellenbosch, South Africa}
 
 \author{D.J. Mar\'{i}n-L\'{a}mbarri}
 \affiliation{iThemba Laboratory for Accelerator Based Sciences, Somerset West 7129, South Africa}
 \affiliation{Department of Physics, University of the Western Cape, P/B X17, Bellville 7535, South Africa}
 \author{K. Lukyanov}
 \affiliation{Joint Institute for Nuclear Research, 141980 Dubna, Russia} 
 \author{N.Y. Kheswa}
 \affiliation{iThemba Laboratory for Accelerator Based Sciences, Somerset West 7129, South Africa}
   \author{R. Neveling}
 \affiliation{iThemba Laboratory for Accelerator Based Sciences, Somerset West 7129, South Africa}
 \author{P. Papka}
 \affiliation{Department of Physics, University of Stellenbosch, Stellenbosch, South Africa}
 
 \author{L. Pellegri}
 \affiliation{iThemba Laboratory for Accelerator Based Sciences, Somerset West 7129, South Africa}
 \affiliation{School of Physics, University of the Witwatersrand, Johannesburg 2050, South Africa}
 \author{V. Pesudo}
 \affiliation{iThemba Laboratory for Accelerator Based Sciences, Somerset West 7129, South Africa}
 \affiliation{Department of Physics, University of the Western Cape, P/B X17, Bellville 7535, South Africa}
 \author{L.C. Pool}
 \affiliation{iThemba Laboratory for Accelerator Based Sciences, Somerset West 7129, South Africa}
 \author{G. Riczu}
 \affiliation{Institute for Nuclear Research, Hungarian Academy of Sciences, Debrecen, Pf. 51, 4001, Hungary-4001}
 \author{F.D. Smit}
 \affiliation{iThemba Laboratory for Accelerator Based Sciences, Somerset West 7129, South Africa}
 \author{J.J. van Zyl}
 \affiliation{Department of Physics, University of Stellenbosch, Stellenbosch, South Africa}
 \author{E. Zemlyanaya}
 \affiliation{Joint Institute for Nuclear Research, 141980 Dubna, Russia}
 
 \date{\today}
 
\begin{abstract}
\begin{description}

\item[Background] Aspects of the nuclear structure of light $\alpha$-conjugate nuclei have long been associated with nuclear clustering based on $\alpha$ particles and heavier $\alpha$-conjugate systems such as $^{12}$C and $^{16}$O. Such structures are associated with strong deformation corresponding to superdeformed or even hyperdeformed bands. Superdeformed bands have been identified in $^{40}$Ca and neighbouring nuclei and find good description within shell model, mean-field and $\alpha$-cluster models. The utility of the $\alpha$-cluster description may be probed further by extending such studies to more challenging cases comprising lighter $\alpha$-conjugate nuclei such as $^{24}$Mg, $^{28}$Si and $^{32}$S.

\item[Purpose] As with many $\alpha$-conjugate nuclei, $^{28}$Si has been suggested to have a number of exotic configurations built on $\alpha$-cluster configurations with large associated deformation. The identification of such bands is challenging as they are high in excitation energy and the in-band transitions are weak compared to the higher-energy out-of-band transitions or particle decay branches. Despite these experimental challenges, recent theoretical and experimental work has pointed to candidate members of a superdeformed band in $^{28}$Si but not its 0$^{+}$ band-head while experimental evidence in support of other predicted configurations is extremely limited. This suggests the value of complementary experimental techniques such as inelastic $\alpha$-particle scattering which naturally favours the population and ready identification of excited 0$^{+}$ states. 

\item[Methods] $\alpha$-particle inelastic scattering from a $^{nat}$Si target at very forward angles including 0\textdegree\ has been performed at the iThemba Laboratory for Accelerator-Based Sciences in South Africa. Scattered particles corresponding to the excitation energy region of 6 to 14 MeV were momentum-analysed in the K600 magnetic spectrometer and detected at the focal plane using two multiwire drift chambers and two plastic scintillators. 

\item[Results] Several $0^+$ states have been identified above 9 MeV in $^{28}$Si. A newly identified 9.71~MeV 0$^{+}$ state is a strong candidate for the band-head of the previously discussed superdeformed band. The multichannel dynamical symmetry of the semi-microscopic algebraic model predicts the spectrum of the excited $0^+$ states. The theoretical prediction is in good agreement with the experimental finding, supporting the assignment of the 9.71-MeV state as the band-head of a superdeformed band.

\item[Conclusion] Excited isoscalar $0^+$ states in $^{28}$Si have been identified. The number of states observed in the present experiment shows good agreement with the prediction of the multichannel dynamical symmetry.

\end{description}
\end{abstract}

\maketitle

\section{Introduction}

In light $\alpha$-conjugate nuclei, some aspects of nuclear structure such as rotational bands comprised of states with large $\alpha$-particle decay widths and large deformations suggest that these nuclei contain clusters of $\alpha$-particles or heavier $\alpha$-conjugate systems such as $^8$Be, $^{12}$C or $^{16}$O. The associated superdeformed rotational bands resulting from these states have been observed in some nuclei such as $^{40}$Ca \cite{PhysRevLett.87.222501} and $^{36}$Ar \cite{PhysRevLett.85.2693} and have received theoretical treatments in the shell model, and mean-field and $\alpha$-cluster models.

In lighter sd-shell nuclei such as $^{24}$Mg, $^{28}$Si and $^{32}$S, the superdeformed rotational bands are less clear. In these cases, the superdeformed bands are expected to lie at a much higher excitation energy than for the previously identified cases. The resulting competition from particle decays and high-energy out-of-band $\gamma$-ray transitions over low-energy in-band transitions makes clear identification of the superdeformed band challenging, especially for low-spin states \cite{PhysRevC.86.064308}. This is because the standard technique used for the identification of superdeformed bands is heavy-ion fusion-evaporation populating high-spin states followed by observation of the resulting $\gamma$ rays from the decays down the rotational band.

An alternative approach to identifying low-spin members of superdeformed bands lies in using $\alpha$-particle inelastic scattering at very forward angles. This approach makes identification of the $0^+$ band-heads of cluster configurations simpler: $\alpha$-particle inelastic scattering preferentially populates low-spin, isoscalar, natural-parity states. Furthermore the differential cross section gives a clear signature of the spin-parity of the populated state. Of course, $\alpha$-particle inelastic scattering cannot be used to probe the superdeformed band at high spin; it is rather a complementary probe to the heavy-ion reactions used to probe the high-spin states.

$^{28}$Si is expected to have a number of different exotic configurations comprised of $\alpha$-conjugate sub-units in addition to strongly deformed mean-field configurations. These include $^{24}$Mg$+\alpha$, $^{12}$C$+^{16}$O and $^{20}$Ne$+2\alpha$ or $^{20}$Ne$+^8$Be configurations \cite{PhysRevC.80.044316,PhysRevC.86.064309,Kimura_private_comm}. However, despite the considerable theoretical investigation there is a lack of experimental evidence for these configurations. Although there is strong theoretical and experimental evidence for a superdeformed band in $^{28}$Si \cite{PhysRevC.86.064309,Cseh2016312,PhysRevC.86.064308}, the location of the $0^+$ band-head remains unknown. The locations of other excited $0^+$ states in $^{28}$Si are unclear with some contradictory spin-parity assignments made for some states \cite{BrenneisenIII}.

The theoretical description  of the high-lying cluster states populated in reactions using $\alpha$ particles and other light heavy-ion beams is a difficult theoretical task. Fully microscopic models can address only special states such as the superdeformed one due to the obvious computational difficulties. Therefore, predictions of the detailed spectrum using fully microscopic models are not available. Phenomenological models on the other hand usually have unwanted ambiguities and so the correspondence between the experimental observation and the theoretical description is not well established.

Here, in order to describe the $^{28}$Si $0^+$ states, a semi-microscopic approach is applied \cite{Cseh2016312}, based on the multichannel dynamical symmetry (MUSY) which connects the shell and cluster models \cite{PhysRevC.50.2240}. It provides a unified multiplet-structure of the two models, applying the same Hamiltonian. Thus the relationship between the experimental and the model spectra is established in the ground-state region where there is no ambiguity, and making extrapolation to higher energies possible.

In this paper, we report a study of the inelastic scattering $\alpha$-particles from a silicon target at scattering angles of between 0\textdegree\ and 6\textdegree\ to locate $0^+$ states in $^{28}$Si and compare the experimental data to state-of-the-art semi-microscopic MUSY calculations.

\section{Experiment}

A 200-MeV beam of $\alpha$ particles was transported down a dispersion-matched beamline and was incident upon a 230-$\mu$g/cm$^{2}$ $^{nat}$Si target. Particles resulting from the reactions were momentum-analysed in the K600 QDD magnetic spectrometer \cite{Neveling201129}. Scattered particles were incident upon two vertical drift chambers (VDCs), which measured horizontal and vertical positions at the focal plane, and a 1/4-inch thick  plastic scintillator. Particles were identified by the time between the particle hitting the plastic scintillator and the next RF reference pulse for the cyclotron, corresponding to the time-of-flight of the scattered particle through the spectrometer, as well as the energy deposited within the plastic scintillator.

In the 0\textdegree\ scattering experiment, the circular spectrometer aperture covered $\theta_{lab}<2$\textdegree. In this mode, unscattered beam was transported through the spectrometer, past the high-momentum side of the focal plane and was stopped by a Faraday cup located within the wall of the experimental vault. In the 0\textdegree\ mode, there was a flat featureless background resulting from target-induced Coulomb scattering. In order to be able to subtract the background resulting from this scattering, the spectrometer was operated in focus mode: the quadrupole located just after the aperture into the spectrometer was used to focus reaction products to a vertically narrow band on the focal plane. 

In the small-angle scattering experiment the centre of the spectrometer aperture was placed at a scattering angle of 4\textdegree, covering scattering angles from 2\textdegree\ to 6\textdegree. In this mode, the unscattered beam was stopped in a Faraday cup adjacent to the aperture into the spectrometer at the spectrometer quadrupole. The background from target-induced Coulomb scattering was much lower, and so the background correction used for 0\textdegree\ data was no longer essential. Thus the spectrometer was operated in under-focus mode: the quadrupole at the entrance of the spectrometer was weakened so that scattered particles are focussed less in the vertical direction than in the focus mode. This allowed scattering angles to be calculated from the horizontal trajectory and vertical position with which scattered particle traversed the focal plane \cite{Neveling201129}.

\section{Data Analysis}
\label{sec:DA}

\begin{figure*}[htp]
\includegraphics[width=\textwidth]{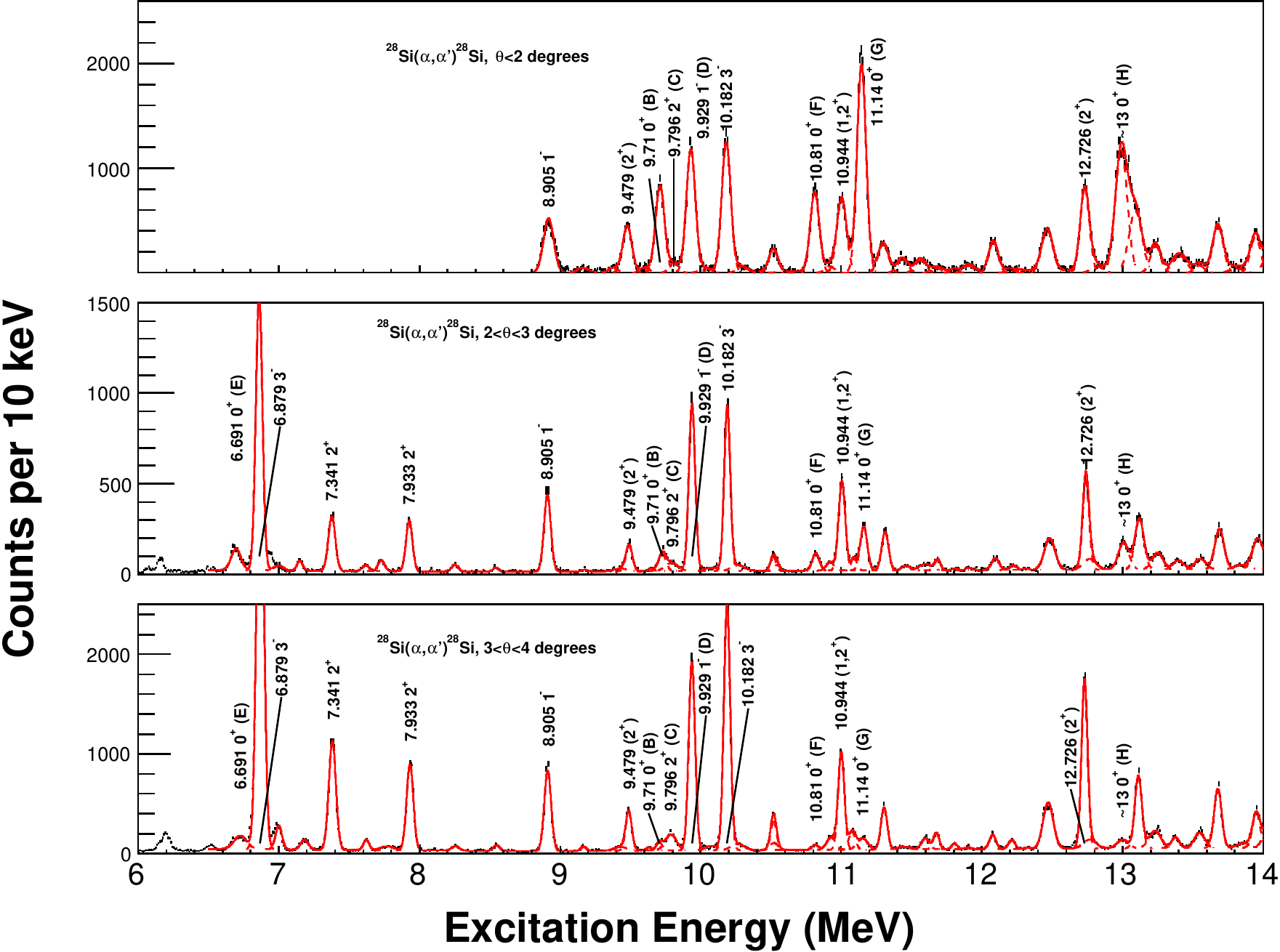}
 \caption{(Top) Background-subtracted $^{28}$Si($\alpha,\alpha^\prime$)$^{28}$Si spectrum at 0\textdegree\ with combined fit (solid line). (Middle) $^{28}$Si($\alpha,\alpha^\prime$)$^{28}$Si spectrum for the 2\textdegree-3\textdegree\  angle bite with combined fit (solid line). (Bottom) $^{28}$Si($\alpha,\alpha^\prime$)$^{28}$Si spectrum for the 3\textdegree-4\textdegree\ angle bite with combined fit (solid line). Prominent states with known energies and spin-parities have been identified in the spectra.}
 \label{fig:spectrum_0deg}
\end{figure*}

The analysis of 0\textdegree\ K600 data has been described in detail elsewhere \cite{Neveling201129} and only the main points are summarised here. Scattered $\alpha$ particles were identified based on the time-of-flight of the particles through the spectrometer and the energy deposited in the scintillator at the focal plane. To optimise the position resolution of the focal plane, the horizontal position was corrected according to the horizontal angle at the focal plane and the vertical focal plane position. Spectra were then rigidity-calibrated using known states in $^{24}$Mg and $^{28}$Si on a run-by-run basis to account for any small shifts in the accelerator fields. To account for the target-induced Coulomb background a well-established technique \cite{Neveling201129,Tamii2009326} was employed: two background spectra were constructed from off-focus sections of the focal plane and then subtracted from the in-focus spectrum. The resulting background-subtracted excitation energy spectrum is shown in the top 
panel of 
Figure \ref{fig:spectrum_0deg}. The energy resolution obtained was 80 keV, FWHM.

In the small-angle experiment, after the corrections for the focal plane angle, the vertical focal plane position and field shifts, the resolution was 65 keV, FWHM. The small-angle experiment was run on a different weekend from the 0\textdegree\ experiment and differences in the set-up of the dispersion-matched beam account for the difference in final energy resolution.

\begin{figure*}
\includegraphics[width=\textwidth]{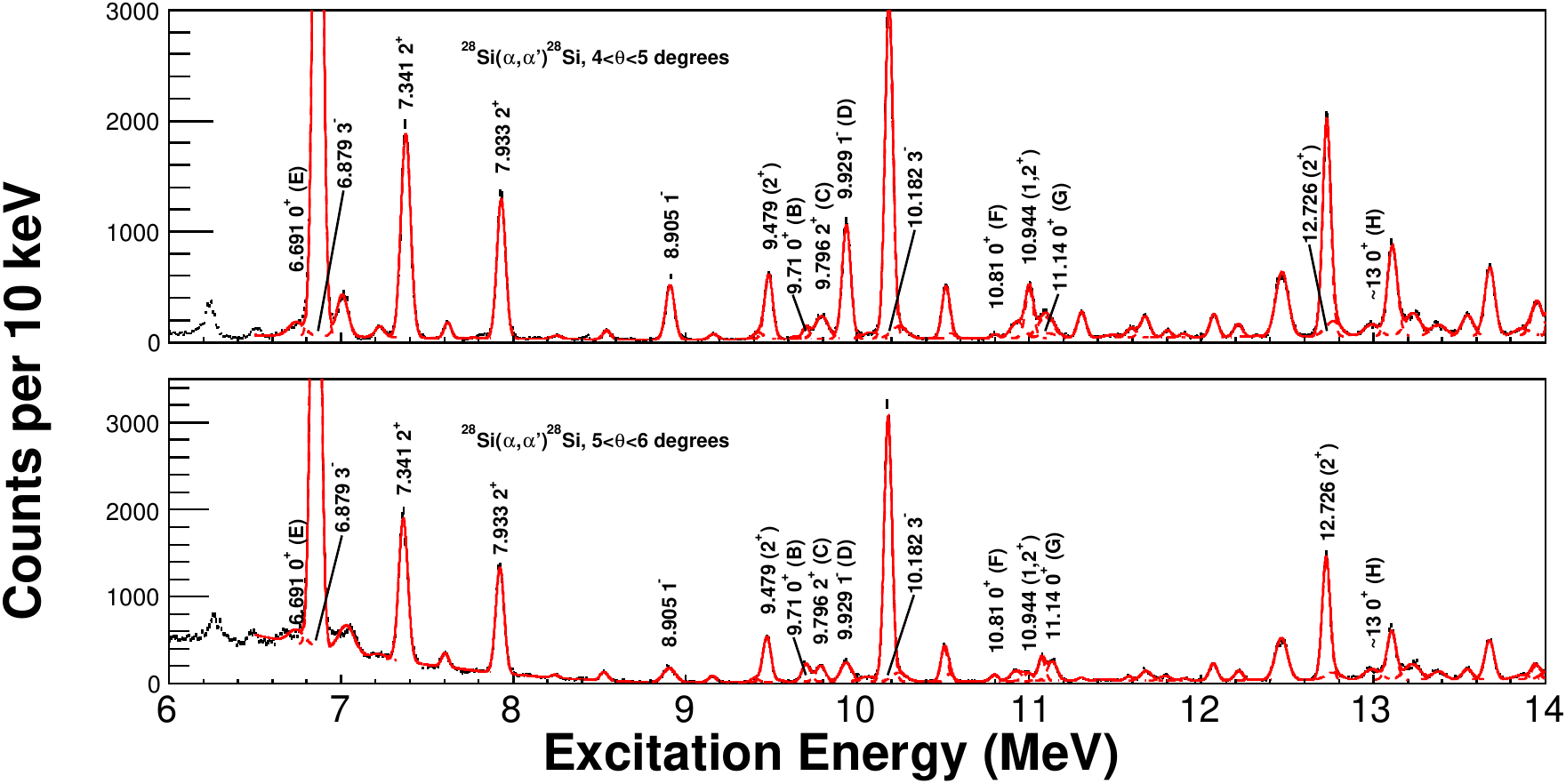}
 \caption{Spectra for the (top) 4-5 degree and (bottom) 5-6 degree angle ranges with combined fit.}
 \label{fig:spectrum_high_angles}
\end{figure*}

The spectra for $^{28}$Si($\alpha$,$\alpha^\prime$)$^{28}$Si reactions for angle ranges of 2\textdegree-3\textdegree\ and 3\textdegree-4\textdegree\ are shown in the middle and bottom panels, respectively, of Figure \ref{fig:spectrum_0deg}. The spectra for the angle ranges 4\textdegree-5\textdegree\ and 5\textdegree-6\textdegree\ are shown in the top and bottom panels, respectively, of Figure \ref{fig:spectrum_high_angles}.

The excitation energy spectra for different angle bins were fitted with a function composed of a Gaussian peak for each experimental state and a linear background which accounted for the continuum and experimental background. An additional phenomenological quadratic background component is added at $E_{x}<9$ MeV to account for scattering from hydrogen. The minimum width of these Gaussians was determined by the experimental resolution taken from the strong $1^-$ state at 9.929 MeV.

To calculate the differential cross section, the efficiency of the focal plane is required. This is the product of the efficiencies for each wire plane. To get the efficiency of one wire plane, the ratio of the number of events which are acceptable \cite{Neveling201129} in all the wire planes is compared to the number of events which are acceptable for all of the wire planes except the wireplane for which the efficiency is being calculated \cite{Hiro_Internal_Memo}. For example, the efficiency of the X1 wire plane, $\eta_{X1}$, is given by:
\begin{equation}
 \eta_{X1} = \frac{\text{Events acceptable in X1, X2, U1 and U2}}{\text{Events acceptable in X2, U1 and U2}}.
\end{equation}
The overall efficiency of the focal plane is 67\%.

% {\it Target contamination: The use of a $^{nat}$Si target makes it possible that some of the states observed result from the other stable silicon isotopes ($^{29}$Si and $^{30}$Si) as well as from common target contaminants such as $^{12}$C or $^{16}$O. Data were taken at small angles with the elastic scattering peak and the low-lying excitation region on the focal plane. [I have a figure of this that I can include if required - not sure how to talk about things here.] The new $0^+$ states observed in the present experiment}

\section{Results and spin assignments of states}

\begin{figure}
\includegraphics[width=0.475\textwidth]{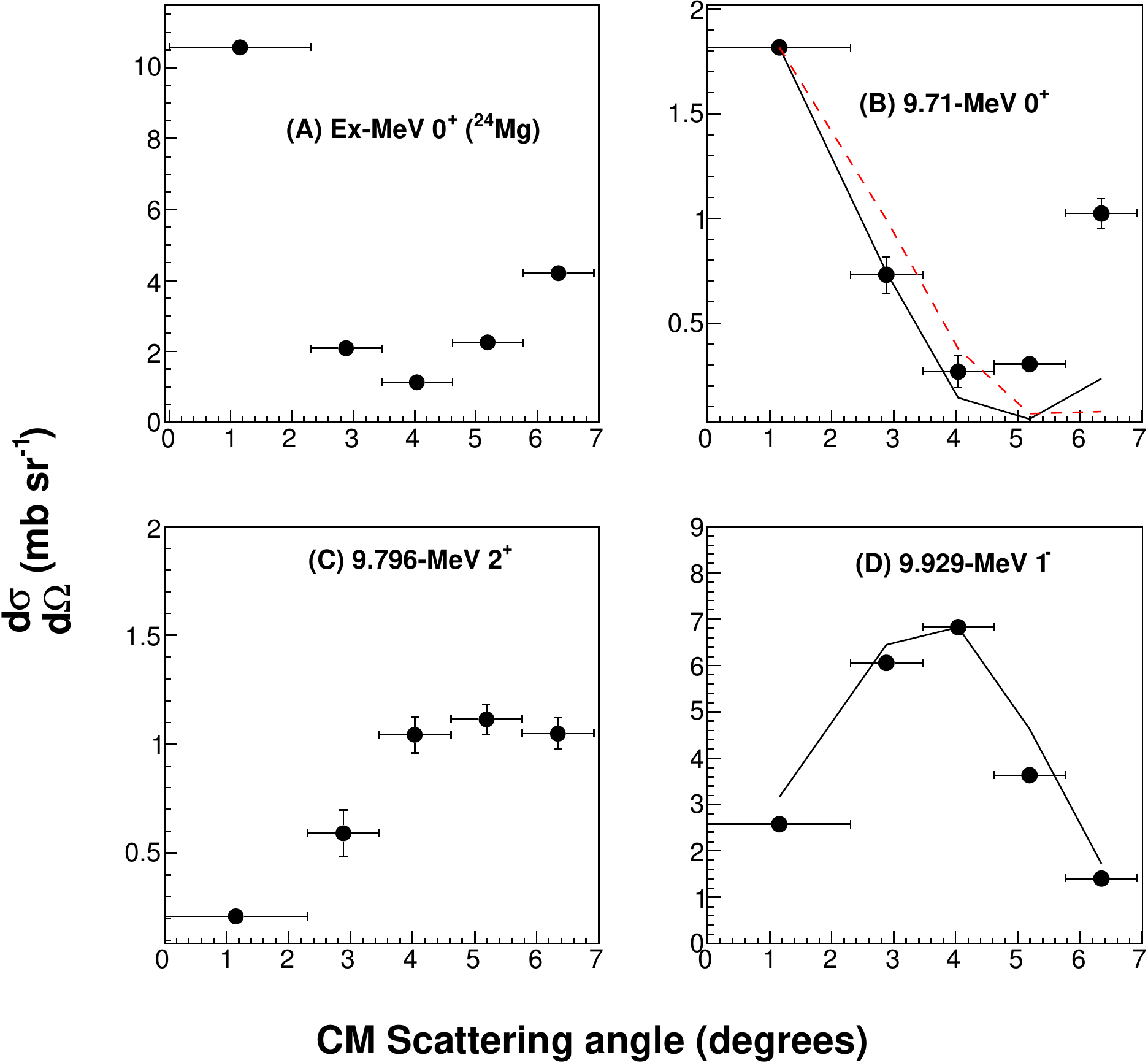}
 \caption{Differential cross sections for states (A) 9.305-MeV $0^+$ state in $^{24}$Mg, (B-D) states in $^{28}$Si. The energies and $J^\pi$ assignments of the state A, C and D are taken from literature \protect\cite{ENSDF}. State B is new. The angle uncertainty for each point corresponds to the angle bite covered. DWBA curves calculated with parameter set II (solid line) averaged over the angle ranges are plotted for $\ell=0$ and $\ell=1$ states - for the 9.71-MeV $0^+$ state the DWBA curve for parameter set I is also plotted (dashed line).}
 \label{fig:dsig_dOmg}
\end{figure}

\begin{figure}
 \includegraphics[width=0.475\textwidth]{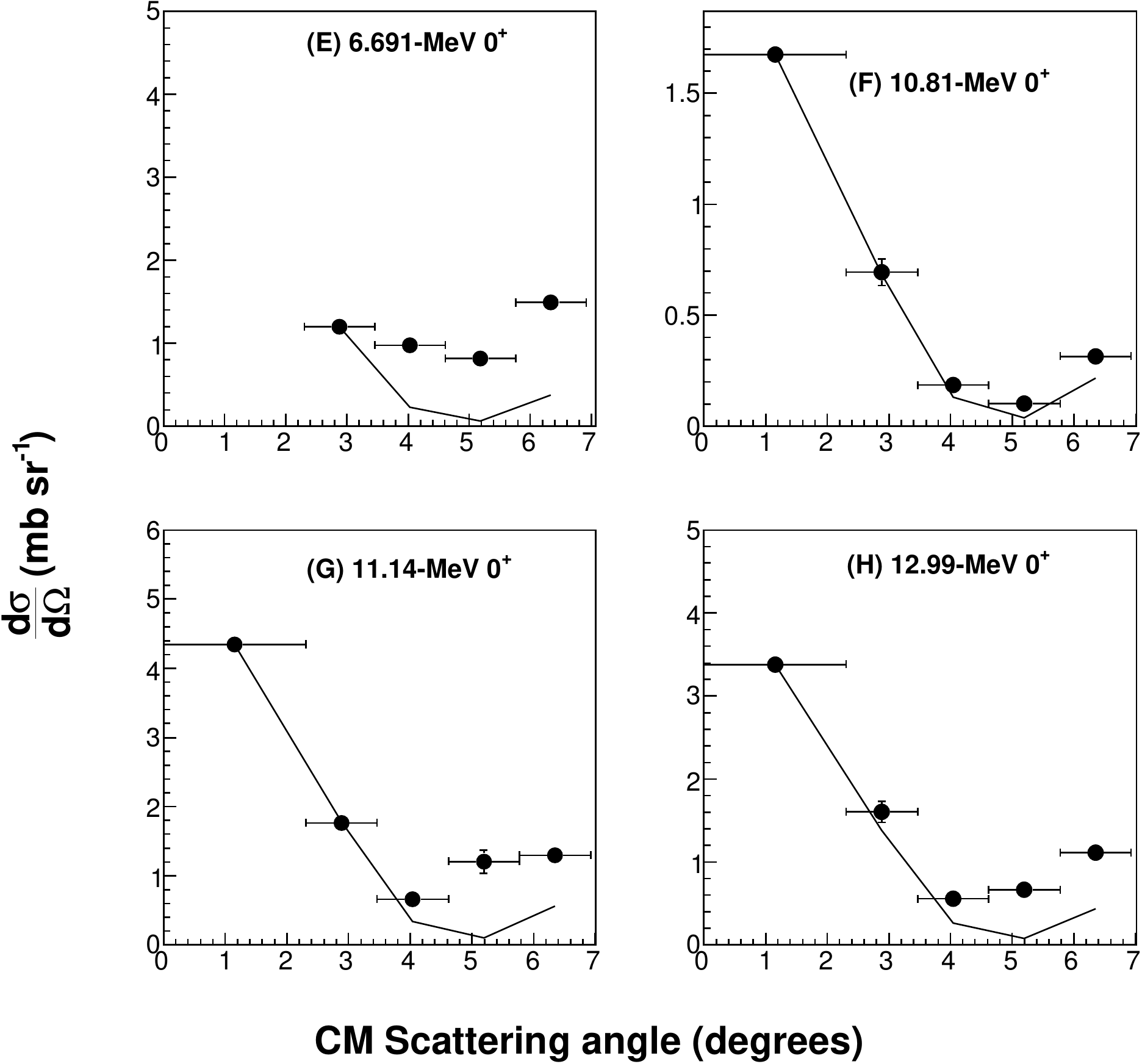}
\caption{As Figure \ref{fig:dsig_dOmg} for the additional $0^+$ states in $^{28}$Si. The 6.691-MeV state is the $0^+_3$ state in $^{28}$Si and does not fall on the focal plane for the 0\textdegree\ data. Table \ref{tab:Si28_results} gives more details as to the properties of the states shown.}
\label{fig:dsig_dOmg2}
\end{figure}

Extracted differential cross sections for selected states are presented in Figures \ref{fig:dsig_dOmg} and \ref{fig:dsig_dOmg2}. A summary of the properties of observed states is given in Table \ref{tab:Si28_results}. To extract the $\ell$-values, the differential cross sections are compared to DWBA calculations. The optical model potential came from a folding potential \cite{Lukyanov2006} which was then fitted to extract a potential of Woods-Saxon form. The reduced radius and the diffusivity of the potential were adjusted to better reproduce the experimentally observed differential cross sections - the initial parameters are given as `Set I' and final values are given as `Set II' in Table \ref{tab:Si28_OMPs}. The DWBA curves were then averaged over the appropriate angle bite and the resulting prediction for the differential cross section was calculated. The DWBA points for $\ell=0$ transitions were normalised to the differential cross section for the $\theta_{lab}<2$\textdegree\ datum. For $\ell=1$ transitions, 
the DWBA curves were normalised to the 3\textdegree$<\theta_{lab}<$4\textdegree\ datum

\begin{table}
 \begin{tabular}{c | c | c | c}
  $E_x$ / MeV & $E_x$ / MeV \protect\cite{ENSDF} & $J^\pi$ & $\frac{d\sigma}{d\Omega}(\theta_{lab}<2)$\footnote{For $J^\pi=0^+$ states only.} / mb/sr \\
  \hline \hline
  9.305\footnote{This is a state in $^{24}$Mg used for comparison.} & 9.30539(24) & $0^+$ & 14.08(28) \\
  9.71(2) & N/A & $0^+$ & 2.42(5) \\
  9.81(3) & 9.79595(14) & $2^+$ & N/A\\
  9.93(1) & 9.9292(17) & $1^-$ & N/A\\
  6.69(5) & 6.69074(15) & $0^+$ & N/A\footnote{Not on focal plane in 0\textdegree\ mode.} \\
  10.81(3) & 10.8055(10) & $0^+$\footnote{This state is assigned as $J^\pi=2^+$ in Ref. \protect\cite{ENSDF}. We argue that that $J^\pi$ assignment is incorrect. See the text for more details.} & 2.23(4)\\
  11.14(2) & 11.142(1)\footnote{This state is one of two observed around 11.14 MeV in Ref. \protect\cite{BrenneisenIII} which have erroneously combined by the compiler \cite{Endt19901,ENSDF}. See the text for more details.} & $0^+$ & 5.79(7) \\
  12.99(2)\footnote{Probably an unresolved doublet.} & \parbox{2cm}{12.976(2)\\ 13.0398(5)} & $0^+$ & 4.51(11) \\
 \end{tabular}
  \caption{Details of the states observed in the present experiment. The differential cross section measured in the 0\textdegree\ experiment is provided for $0^+$ states.}
 \label{tab:Si28_results}
\end{table}

\begin{table}
 \begin{tabular}{ c | c | c }
 Parameter & Set I & Set II \\
 \hline \hline
 $E_\alpha$ / MeV & 200 & 200\\
 $V_R$ / MeV & -95.52 & -95.52 \\
 $r_R$ / fm  & 1.05 & 1.35 \\
 $a_R$ / fm  & 0.99 & 0.65 \\
 $V_I$ / MeV & -68.00 & -68.00 \\
 $r_I$ / fm  & 0.94 & 1.35  \\
 $a_I$ / fm  & 0.79 & 0.65 \\
 \end{tabular}
\caption{Optical model parameters for $^{28}$Si($\alpha,\alpha^\prime$)$^{28}$Si reactions. Set I are the OMP parameters from Ref. \protect\cite{Lukyanov2006}. Set II are the OMP parameters with modified radial terms which better reproduce the data in the present experiment.}
\label{tab:Si28_OMPs}
\end{table}

In addition to the DWBA calculations, we used states with known spin-parities to test the behaviour of the differential cross sections: in the absence of another well-known $0^+$ state in $^{28}$Si within our excitation energy bite at 0\textdegree, we use the 9.305-MeV $0^+_6$ state in $^{24}$Mg (Figure \ref{fig:dsig_dOmg}(a)). We also observe a weak 0$^{+}$ state at 12.085(15) MeV which is the 12.049(2)-MeV $0^+$ state in $^{16}$O originating from the water contamination on the target. The excitation energy is shifted due to the differing masses of $^{16}$O and $^{28}$Si.

It is clear from the differential cross sections shown in Figure \ref{fig:dsig_dOmg} that $\ell=0$ and $\ell=1$ transitions exhibit particular angular distributions, with the $\ell=0$ transition showing a strong maximum at 0\textdegree\ and a minimum at around 4\textdegree, while the $\ell=1$ transition has a maximum around 4\textdegree\ and falls off at higher angles. As $0^+$ states were the focus of this experimental study, scattering angles greater than 6\textdegree\ were not measured meaning that only $\ell=0$ and $\ell=1$ transitions may be firmly assigned.

For some of the $0^+$ states, the DWBA curves do not reproduce the increase in the differential cross section at higher angles. This is likely results from additional states with higher spins at around the same excitation energy or from multistep contributions to the cross section which are not accounted for in the DWBA calculations. In no case does it affect the assignment of a $0^+$ state as this is based on the differential cross section reaching a maximum at 0\textdegree.

All of the states discussed in this section are strongly populated in the $(\alpha,\alpha^\prime)$ reaction. If the states resulted from target contaminants such as $^{29}$Si or $^{30}$Si, the cross sections for these states would have to be extremely high (e.g. $\sigma(\theta_{lab}<2) \sim 70$ mb for the 9.71-MeV state) to match the observed experimental yield, assuming natural isotopic abundance in the target. This is around five times higher than comparable cross sections in nearby nuclei (see, as an example, the cross section for the 9.305-MeV $0^+$ state in $^{24}$Mg in Figure \ref{fig:dsig_dOmg}) and leads us to conlude that it is unlikely that any of the states listed in Table \ref{tab:Si28_results} result from $^{29}$Si or $^{30}$Si. The number and position of narrow $0^+$ states in $^{12}$C and $^{16}$O below 14 MeV are well known and have been previously observed using the same reaction at the same facility \cite{2016arXiv161007437L,KCWLThesis} and do not match the observed states (with the exception of the 12.05-MeV state from $^{16}$O which is identified due to its shift on the focal plane).

The $0^+_3$ state at 6.691 MeV (state E) is observed but only at higher angles and in a focal plane region with a strong background due to scattering from protons in the target. The state is only observed at higher angles because the focal plane excitation energy bite only extends down to just below 9 MeV in the 0\textdegree\ experiment. The behaviour of the observed data is consistent with the trend of other $0^+$ states observed in the present experiment.

The state observed at 9.71 MeV is newly observed in the present experiment and is unambiguously assigned as $J^\pi = 0^+$. Extrapolating the candidate superdeformed band in $^{28}$Si from Refs. \cite{PhysRevC.86.064308,BrenneisenIII} suggests that the band-head should lie at around 9.3 MeV. The 9.71-MeV state is the only $0^+$ state in this excitation energy region (from around 8.8 MeV to 10 MeV) and is the only observed candidate for the band-head.

Another $0^+$ state is observed at 10.81 MeV. There is a state listed at 10.806 MeV with $J^\pi=2^+$ \cite{ENSDF} which has been observed in $^{28}$Si($e,e^\prime$)$^{28}$Si \cite{Schneider197913} and $^{27}$Al($p,\gamma$)$^{28}$Si reactions \cite{BrenneisenI}. In the latter case, it is populated through decay of the 13.321-MeV $T=1,1^+$ state. While we cannot exclude the possibility that there are two near-degenerate states here, the previously observed gamma decay would be consistent with the existence of a single 0$^{+}$, $T=0$ state where the decay was a strong isovector M1 transition. The present measurement is more selective in terms of 0$^{+}$ assignments and, on the balance of probability, there is likely a single state with $J^\pi=0^+$ at this energy.
 
The $0^+$ state at 11.14 MeV has been previously observed in the $^{24}$Mg($\alpha,\gamma$)$^{28}$Si reaction by Brenneisen \textit{et al.} with an energy of 11.142 MeV \cite{BrenneisenI,BrenneisenII,BrenneisenIII}. Brenneisen \textit{et al.} also observed a $2^+$ state at 11.148 MeV. However, the compiler \cite{ENSDF} has suggested that these states are the same. From the current experiment in which a strong $0^+$ state is observed, we conclude that there are two states, one $2^+$ state at 11.148 MeV and one $0^+$ state at 11.142 MeV. Based on the large resonance strength observed in the $^{24}$Mg($\alpha,\gamma$)$^{28}$Si reaction \cite{PhysRevC.77.055801}, it is probable that at least one of these states is a $^{24}$Mg$+\alpha$ cluster state. We note that the reaction used in the present experiment strongly populates cluster states \cite{Kawabata20076}, which is suggestive of the $0^{+}$ state having a cluster structure.

There is at least one $0^+$ state at around 13 MeV; previous experimental studies have observed three $0^+$ states in this region (12.976(2), 13.0398(5) and 13.234(2) MeV) \cite{Cseh198243,ENSDF} though one is not isoscalar \cite{ENSDF}. In the present experiment, the 13.234-MeV state is not observed and the 12.976-MeV and 13.0398-MeV states cannot be resolved. We assume the energies of the states given in Ref. \cite{Cseh198243}.

\section{Comparison with a symmetry-based prediction}
 
Here we address the question: what is the spectrum of $0^+$ states of the $^{24}$Mg+$^{4}$He clusterisation in the energy range of the present experimental investigation? When doing so, we apply the Hamiltonian and the multichannel dynamical symmetry of Ref. \cite{Cseh2016312} which was used to descibe the low-energy spectrum in terms of the semi-microscopic algebraic quarted model. This gives a unified description of the shell model and cluster structures of $^{28}$Si. In earlier studies \cite{PhysRevC.50.2240,PhysRevC.87.067301} only different binary clusterisations were considered in this way. Since the model-spaces are constructed microscopically, the resulting spectrum is a pure prediction without any ambiguity or fitting to the experimentally observed energies.

First we briefly summarise the basic ideas of the approach and then we present the spectrum of $0^+$ states.

{\it The  semi-microscopic algebraic quartet model} \cite{Cseh2015213} is a symmetry-governed truncation of the no-core shell model \cite{Dytrych2008} that describes the quartet excitations in a nucleus. A quartet is formed by two protons and two neutrons which interact with each other very strongly as a consequence of the short-range attractive forces between the nucleons inside a nucleus \cite{PhysRevLett.25.1043}. The interaction between the different quartets is weaker. In this approach the L-S coupling is applied, the model space has a spin-isospin sector characterized by Wigner's U$^{ST}$(4) group \cite{PhysRev.51.106}, and a spatial part described by Elliott's U(3) group \cite{Elliott128}. Four nucleons form a quartet \cite{HARVEY1973191} when their
spin-isospin symmetry is \{1,1,1,1\}, and their permutational symmetry is \{4\}. This definition allows two protons and two neutrons to form a
quartet even if they sit in different shells. 

{\it The semi-microscopic algebraic cluster model} \cite{CSEH1992173,CSEH1994165}, as with other cluster models, classifies the relevant degrees of freedom of the nucleus into two categories: they belong either to the internal structure of the clusters or to their relative motion. The internal structure of the clusters is handled in terms of Elliott's shell model \cite{Elliott128} with U$^{ST}$(4)$\otimes$U(3) group structure (as discussed above). The relative motion is taken care of by the vibron model \cite{IachelloJChemPhys_77_3046}, which is an algebraic model of the dipole motion also with a U(3) basis. For a two-cluster-configuration this model has a group-structure of U$^{ST}_{C_1}$(4)$\otimes$U$_{C_1}$(3) $\otimes$ U$^{ST}_{C_2}$(4)$\otimes$U$_{C_2}$(3) $\otimes$ U$_R$(4). In this case the model space is also constructed microscopically, i.e. the Pauli-forbidden states are excluded.

{\it The multichannel dynamical symmetry} \cite{PhysRevC.50.2240,PhysRevC.87.067301} connects different cluster configurations (including the shell model limit) in a nucleus. Here the word channel refers to the reaction channel that defines the cluster configuration.
 
The MUSY is a composite symmetry of a composite system. The system has two (or more) different clusterisations, each of them having dynamical symmetries which are connected to each other by the symmetry of the pseudo space of the particle indices that change from one configuration to the other.
               
When the multichannel dynamical symmetry holds, then the spectra of different clusterisations are related to one another by very strong constraints. The MUSY provides a unified multiplet structure of different cluster configurations in which the corresponding energies and $E2$ transitions coincide exactly. Of course, it cannot be decided {\it a priori} whether the MUSY holds, rather one assumes the symmetry and compares its results with the experimental data.

{\it The distribution of $0^+$ states.} $^{28}$Si has a well-established band-structure, and the SU(3) quantum numbers of several bands could be assigned as a joint conclusion of experimental and theoretical investigations \cite{Sheline1982263}. In Ref. \cite{Cseh2016312} its energy spectrum was calculated within the SAQM approach by fitting the parameters to the well-known states. A U(3) dynamically symmetric Hamiltonian was applied, which is invariant with respect to the transformation between the quartet and cluster model. It is expressed in terms of the invariant operators of the group-chain: 
U(3) $\supset$ SU(3)  $\supset$ SO(3):
\begin{equation}
{\hat H} = (\hbar \omega) {\hat n}  + a{\hat C}^{(2)}_{SU3} + b{\hat C}^{(3)}_{SU3} + d \frac{1}{2\theta}{\hat L}^2.
\end{equation}
The first term is the harmonic oscillator Hamiltonian (linear invariant of the U(3)), with a strength obtained from the systematics \cite{BLOMQVIST1968545}
$\hbar \omega$ = $45 A^{-\frac{1}{3}} -25  A^{-\frac{2}{3}}$ MeV = 12.11 MeV.
%($n$  is the number of excitation quanta compared to the ground state.)
The second-order invariant of the SU(3) (${\hat C}^{(2)}_{SU3}$) represents the quadrupole-quadrupole interaction, while the third-order Casimir-operator (${\hat C}^{(3)}_{SU3}$) distinguishes between the prolate and oblate shapes. The moment of inertia, $\theta$ is calculated classically for the rigid shape
determined by the U(3) quantum numbers (for a rotor with axial symmetry) \cite{Cseh2015213}. The $a$, $b$ and  $d$ parameters were fitted to the low-lying experimental states: $ a = -0.133$ MeV, $b =0.000444$ MeV $d = 1.003$ MeV.

\begin{figure}
\includegraphics[width=0.45\textwidth]{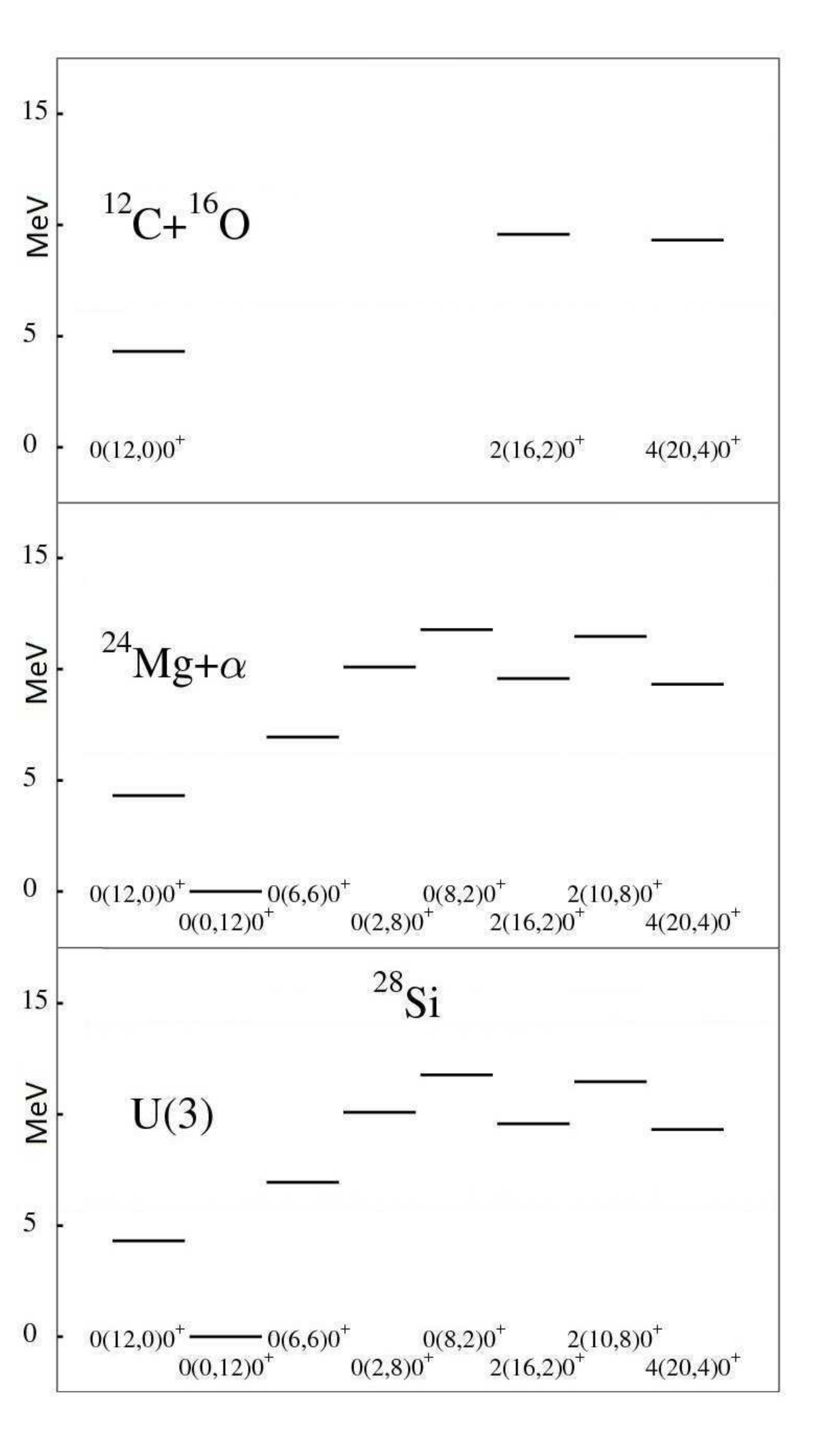}
\caption{The spectra of the $0^+$ states in the $^{28}$Si nucleus and in its $^{24}$Mg+$^{4}$He and $^{16}$O+$^{12}$C clusterisations, as predicted by the Hamiltonian of the multichannel dynamical symmetry \protect\cite{Cseh2016312}. The states are characterised by the $n(\lambda,\mu)K^\pi$ quantum numbers, where $n$ is the major shell excitation, and $(\lambda,\,u)$ refers to the SU(3) representation i.e. the quadrupole deformation.
\label{fig:spectrum1}}
\end{figure}

\begin{figure}
%[placement !]
\includegraphics[width=0.3\textwidth]{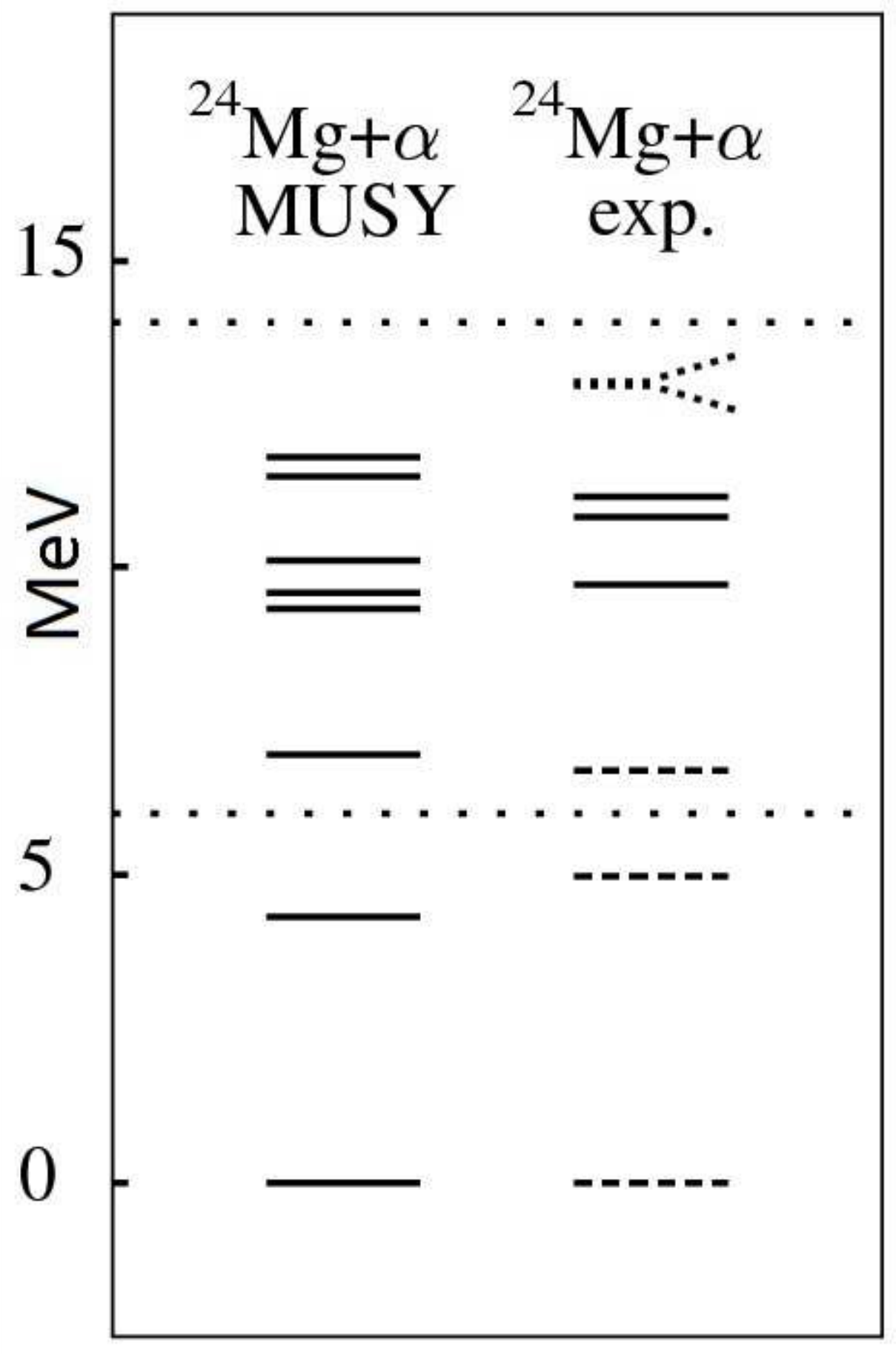}
\caption{Comparison of the $^{24}$Mg+$^{4}$He $0^+$ spectra from the theoretical prediction (of Figure \ref{fig:spectrum1}) and the experimental observation. The energy-window of the present experiment is indicated by the dotted lines in both panels. In the experimental part the solid lines show the observed resonances of the present work, the dotted Y-shaped lines are the states from the $^{24}$Mg+$^{4}$He  measurements \protect\cite{Cseh198243} which are not resolved in the present experiment. The dashed lines correspond to known low-lying $0^+$ states which were not measured at 0\textdegree\ in the present experiment.
\label{fig:spectrum2}}
\end{figure}

Here this Hamiltonian is used for the calculation of $0^+$ states in the quartet spectrum as well as in the $^{24}$Mg+$^{4}$He and $^{16}$O+$^{12}$C cluster spectra. The result is shown in Figure \ref{fig:spectrum1}. The spectrum of $0^+$ states predicted by the MUSY for the $^{24}$Mg+$^{4}$He system is compared with the experimental observation in Figure \ref{fig:spectrum2}. We find the agreement remarkable. 

The theoretical state corresponding to the 9.71-MeV 0+ state has a 4(20,4)0$^+$ structure: a 4p-4h excitation with $\beta = 0.88$ and $\gamma = 9$\textdegree. This provides a strong theoretical justification for interpreting this state as the band-head for the superdeformed band. For comparison, we recall that the deformed rotational bands in $^{36}$Ar and $^{40}$Ca have been described in terms of the shell model as 4p-4h and 8p-8h excitations from the sd-shell into the fp-shell \cite{PhysRevLett.85.2693,PhysRevLett.87.222501}.

The energy of the candidate state is around 400 keV higher than expected from a simple linear extrapolation from the known members of the superdeformed band. However, in light nuclei deviations from linearity in rotational bands are not uncommon (see e.g. Refs. \cite{PhysRevC.90.024327,PhysRevLett.87.142502} or the normally deformed band in $^{28}$Si \cite{PhysRevC.86.064308}). It is possible that the raising in energy of the band-head is due to mixing between different $0^+$ states which cause the ground state energy to be slightly lowered and the corresponding excited states to be pushed slightly higher in energy. This mixing effect is unlikely to result in any change in the moment of inertia of the rotational band.

\section{Conclusions}

Excited isoscalar states in the self-conjugate nucleus $^{28}$Si have been observed using $\alpha$-particle inelastic scattering. The strong selectivity of this reaction to low-spin natural-parity isoscalar transitions allows a possible band-head of the candidate superdeformed band in $^{28}$Si to be identified at 9.71 MeV. Experimental confirmation that this state lies within the candidate superdeformed band is not possible in this experiment though the theoretical model supports this state being assigned as a member of the superdeformed band.

A $0^+$ state has been observed at 10.81 MeV. No structural assignments are yet possible for this state.

A $0^+$ state has been identified at 11.14 MeV. We suggest that this state is the 11.142-MeV state observed in previous experiments \cite{BrenneisenI,BrenneisenII,BrenneisenIII}, and that there is a separate $2^+$ state at a similar excitation energy. Furthermore, it is probable, based on the results of $^{24}$Mg($\alpha,\gamma$)$^{28}$Si reactions that at least one of the states around 11.14 MeV is a cluster state.

A concentration of monopole strength was observed around 13 MeV. However, based on previous experimental studies of $^{28}$Si, this strength corresponds to multiple unresolved $0^+$ states \cite{Cseh198243}.

The multichannel dynamical symmetry of the semi-microscopic algebraic model predicts six states in the excitation region covered in the present experiment, in good agreement with the experimental results which show four distinct states plus at least one further state in the region around 13 MeV \cite{Cseh198243}. The theoretical spectrum is obtained as a parameter-free prediction of the MUSY, without any ambiguity. The Hamiltonian was determined in the quartet-model description of low-lying well-established bands of the $^{28}$Si nucleus \cite{Cseh2016312}. In the $^{12}$C$+^{16}$O channel, the same Hamiltonian gives a detailed spectrum also in good agreement with experimental observations \cite{Cseh2016312}.

\section{Acknowledgements}

The authors thank the accelerator group at iThemba LABS for the provision of the high-quality dispersion-matched beam. In addition, PA thanks M. Kimura of Hokkaido University for useful discussions on the structure of $^{28}$Si. The K600 facility is supported by the NRF. RN acknowledges financial support from the NRF through grant number 85509. This work was supported in part by the Hungarian Scientific Research Fund - OTKA (Grant No. K112962). The studies of SSD were partially supported by the DFNI-T02/19 grant of the Bulgarian Science Foundation.

\bibliography{Si28_zero_plus_states_final}

\begin{thebibliography}{37}%
\makeatletter
\providecommand \@ifxundefined [1]{%
 \@ifx{#1\undefined}
}%
\providecommand \@ifnum [1]{%
 \ifnum #1\expandafter \@firstoftwo
 \else \expandafter \@secondoftwo
 \fi
}%
\providecommand \@ifx [1]{%
 \ifx #1\expandafter \@firstoftwo
 \else \expandafter \@secondoftwo
 \fi
}%
\providecommand \natexlab [1]{#1}%
\providecommand \enquote  [1]{``#1''}%
\providecommand \bibnamefont  [1]{#1}%
\providecommand \bibfnamefont [1]{#1}%
\providecommand \citenamefont [1]{#1}%
\providecommand \href@noop [0]{\@secondoftwo}%
\providecommand \href [0]{\begingroup \@sanitize@url \@href}%
\providecommand \@href[1]{\@@startlink{#1}\@@href}%
\providecommand \@@href[1]{\endgroup#1\@@endlink}%
\providecommand \@sanitize@url [0]{\catcode `\\12\catcode `\$12\catcode
  `\&12\catcode `\#12\catcode `\^12\catcode `\_12\catcode `\%12\relax}%
\providecommand \@@startlink[1]{}%
\providecommand \@@endlink[0]{}%
\providecommand \url  [0]{\begingroup\@sanitize@url \@url }%
\providecommand \@url [1]{\endgroup\@href {#1}{\urlprefix }}%
\providecommand \urlprefix  [0]{URL }%
\providecommand \Eprint [0]{\href }%
\providecommand \doibase [0]{http://dx.doi.org/}%
\providecommand \selectlanguage [0]{\@gobble}%
\providecommand \bibinfo  [0]{\@secondoftwo}%
\providecommand \bibfield  [0]{\@secondoftwo}%
\providecommand \translation [1]{[#1]}%
\providecommand \BibitemOpen [0]{}%
\providecommand \bibitemStop [0]{}%
\providecommand \bibitemNoStop [0]{.\EOS\space}%
\providecommand \EOS [0]{\spacefactor3000\relax}%
\providecommand \BibitemShut  [1]{\csname bibitem#1\endcsname}%
\let\auto@bib@innerbib\@empty
%</preamble>
\bibitem [{\citenamefont {Ideguchi}\ \emph {et~al.}(2001)\citenamefont
  {Ideguchi}, \citenamefont {Sarantites}, \citenamefont {Reviol}, \citenamefont
  {Afanasjev}, \citenamefont {Devlin}, \citenamefont {Baktash}, \citenamefont
  {Janssens}, \citenamefont {Rudolph}, \citenamefont {Axelsson}, \citenamefont
  {Carpenter}, \citenamefont {Galindo-Uribarri}, \citenamefont {LaFosse},
  \citenamefont {Lauritsen}, \citenamefont {Lerma}, \citenamefont {Lister},
  \citenamefont {Reiter}, \citenamefont {Seweryniak}, \citenamefont
  {Weiszflog},\ and\ \citenamefont {Wilson}}]{PhysRevLett.87.222501}%
  \BibitemOpen
  \bibfield  {author} {\bibinfo {author} {\bibfnamefont {E.}~\bibnamefont
  {Ideguchi}}, \bibinfo {author} {\bibfnamefont {D.~G.}\ \bibnamefont
  {Sarantites}}, \bibinfo {author} {\bibfnamefont {W.}~\bibnamefont {Reviol}},
  \bibinfo {author} {\bibfnamefont {A.~V.}\ \bibnamefont {Afanasjev}}, \bibinfo
  {author} {\bibfnamefont {M.}~\bibnamefont {Devlin}}, \bibinfo {author}
  {\bibfnamefont {C.}~\bibnamefont {Baktash}}, \bibinfo {author} {\bibfnamefont
  {R.~V.~F.}\ \bibnamefont {Janssens}}, \bibinfo {author} {\bibfnamefont
  {D.}~\bibnamefont {Rudolph}}, \bibinfo {author} {\bibfnamefont
  {A.}~\bibnamefont {Axelsson}}, \bibinfo {author} {\bibfnamefont {M.~P.}\
  \bibnamefont {Carpenter}}, \bibinfo {author} {\bibfnamefont {A.}~\bibnamefont
  {Galindo-Uribarri}}, \bibinfo {author} {\bibfnamefont {D.~R.}\ \bibnamefont
  {LaFosse}}, \bibinfo {author} {\bibfnamefont {T.}~\bibnamefont {Lauritsen}},
  \bibinfo {author} {\bibfnamefont {F.}~\bibnamefont {Lerma}}, \bibinfo
  {author} {\bibfnamefont {C.~J.}\ \bibnamefont {Lister}}, \bibinfo {author}
  {\bibfnamefont {P.}~\bibnamefont {Reiter}}, \bibinfo {author} {\bibfnamefont
  {D.}~\bibnamefont {Seweryniak}}, \bibinfo {author} {\bibfnamefont
  {M.}~\bibnamefont {Weiszflog}}, \ and\ \bibinfo {author} {\bibfnamefont
  {J.~N.}\ \bibnamefont {Wilson}},\ }\href {\doibase
  10.1103/PhysRevLett.87.222501} {\bibfield  {journal} {\bibinfo  {journal}
  {Phys. Rev. Lett.}\ }\textbf {\bibinfo {volume} {87}},\ \bibinfo {pages}
  {222501} (\bibinfo {year} {2001})}\BibitemShut {NoStop}%
\bibitem [{\citenamefont {Svensson}\ \emph {et~al.}(2000)\citenamefont
  {Svensson}, \citenamefont {Macchiavelli}, \citenamefont {Juodagalvis},
  \citenamefont {Poves}, \citenamefont {Ragnarsson}, \citenamefont {\AA{}berg},
  \citenamefont {Appelbe}, \citenamefont {Austin}, \citenamefont {Baktash},
  \citenamefont {Ball}, \citenamefont {Carpenter}, \citenamefont {Caurier},
  \citenamefont {Clark}, \citenamefont {Cromaz}, \citenamefont {Deleplanque},
  \citenamefont {Diamond}, \citenamefont {Fallon}, \citenamefont {Furlotti},
  \citenamefont {Galindo-Uribarri}, \citenamefont {Janssens}, \citenamefont
  {Lane}, \citenamefont {Lee}, \citenamefont {Lipoglavsek}, \citenamefont
  {Nowacki}, \citenamefont {Paul}, \citenamefont {Radford}, \citenamefont
  {Sarantites}, \citenamefont {Seweryniak}, \citenamefont {Stephens},
  \citenamefont {Tomov}, \citenamefont {Vetter}, \citenamefont {Ward},\ and\
  \citenamefont {Yu}}]{PhysRevLett.85.2693}%
  \BibitemOpen
  \bibfield  {author} {\bibinfo {author} {\bibfnamefont {C.~E.}\ \bibnamefont
  {Svensson}}, \bibinfo {author} {\bibfnamefont {A.~O.}\ \bibnamefont
  {Macchiavelli}}, \bibinfo {author} {\bibfnamefont {A.}~\bibnamefont
  {Juodagalvis}}, \bibinfo {author} {\bibfnamefont {A.}~\bibnamefont {Poves}},
  \bibinfo {author} {\bibfnamefont {I.}~\bibnamefont {Ragnarsson}}, \bibinfo
  {author} {\bibfnamefont {S.}~\bibnamefont {\AA{}berg}}, \bibinfo {author}
  {\bibfnamefont {D.~E.}\ \bibnamefont {Appelbe}}, \bibinfo {author}
  {\bibfnamefont {R.~A.~E.}\ \bibnamefont {Austin}}, \bibinfo {author}
  {\bibfnamefont {C.}~\bibnamefont {Baktash}}, \bibinfo {author} {\bibfnamefont
  {G.~C.}\ \bibnamefont {Ball}}, \bibinfo {author} {\bibfnamefont {M.~P.}\
  \bibnamefont {Carpenter}}, \bibinfo {author} {\bibfnamefont {E.}~\bibnamefont
  {Caurier}}, \bibinfo {author} {\bibfnamefont {R.~M.}\ \bibnamefont {Clark}},
  \bibinfo {author} {\bibfnamefont {M.}~\bibnamefont {Cromaz}}, \bibinfo
  {author} {\bibfnamefont {M.~A.}\ \bibnamefont {Deleplanque}}, \bibinfo
  {author} {\bibfnamefont {R.~M.}\ \bibnamefont {Diamond}}, \bibinfo {author}
  {\bibfnamefont {P.}~\bibnamefont {Fallon}}, \bibinfo {author} {\bibfnamefont
  {M.}~\bibnamefont {Furlotti}}, \bibinfo {author} {\bibfnamefont
  {A.}~\bibnamefont {Galindo-Uribarri}}, \bibinfo {author} {\bibfnamefont
  {R.~V.~F.}\ \bibnamefont {Janssens}}, \bibinfo {author} {\bibfnamefont
  {G.~J.}\ \bibnamefont {Lane}}, \bibinfo {author} {\bibfnamefont {I.~Y.}\
  \bibnamefont {Lee}}, \bibinfo {author} {\bibfnamefont {M.}~\bibnamefont
  {Lipoglavsek}}, \bibinfo {author} {\bibfnamefont {F.}~\bibnamefont
  {Nowacki}}, \bibinfo {author} {\bibfnamefont {S.~D.}\ \bibnamefont {Paul}},
  \bibinfo {author} {\bibfnamefont {D.~C.}\ \bibnamefont {Radford}}, \bibinfo
  {author} {\bibfnamefont {D.~G.}\ \bibnamefont {Sarantites}}, \bibinfo
  {author} {\bibfnamefont {D.}~\bibnamefont {Seweryniak}}, \bibinfo {author}
  {\bibfnamefont {F.~S.}\ \bibnamefont {Stephens}}, \bibinfo {author}
  {\bibfnamefont {V.}~\bibnamefont {Tomov}}, \bibinfo {author} {\bibfnamefont
  {K.}~\bibnamefont {Vetter}}, \bibinfo {author} {\bibfnamefont
  {D.}~\bibnamefont {Ward}}, \ and\ \bibinfo {author} {\bibfnamefont {C.~H.}\
  \bibnamefont {Yu}},\ }\href {\doibase 10.1103/PhysRevLett.85.2693} {\bibfield
   {journal} {\bibinfo  {journal} {Phys. Rev. Lett.}\ }\textbf {\bibinfo
  {volume} {85}},\ \bibinfo {pages} {2693} (\bibinfo {year}
  {2000})}\BibitemShut {NoStop}%
\bibitem [{\citenamefont {Jenkins}\ \emph {et~al.}(2012)\citenamefont
  {Jenkins}, \citenamefont {Lister}, \citenamefont {Carpenter}, \citenamefont
  {Chowdury}, \citenamefont {Hammond}, \citenamefont {Janssens}, \citenamefont
  {Khoo}, \citenamefont {Lauritsen}, \citenamefont {Seweryniak}, \citenamefont
  {Davinson}, \citenamefont {Woods}, \citenamefont {Jokinen}, \citenamefont
  {Penttila}, \citenamefont {Haas},\ and\ \citenamefont
  {Courtin}}]{PhysRevC.86.064308}%
  \BibitemOpen
  \bibfield  {author} {\bibinfo {author} {\bibfnamefont {D.~G.}\ \bibnamefont
  {Jenkins}}, \bibinfo {author} {\bibfnamefont {C.~J.}\ \bibnamefont {Lister}},
  \bibinfo {author} {\bibfnamefont {M.~P.}\ \bibnamefont {Carpenter}}, \bibinfo
  {author} {\bibfnamefont {P.}~\bibnamefont {Chowdury}}, \bibinfo {author}
  {\bibfnamefont {N.~J.}\ \bibnamefont {Hammond}}, \bibinfo {author}
  {\bibfnamefont {R.~V.~F.}\ \bibnamefont {Janssens}}, \bibinfo {author}
  {\bibfnamefont {T.~L.}\ \bibnamefont {Khoo}}, \bibinfo {author}
  {\bibfnamefont {T.}~\bibnamefont {Lauritsen}}, \bibinfo {author}
  {\bibfnamefont {D.}~\bibnamefont {Seweryniak}}, \bibinfo {author}
  {\bibfnamefont {T.}~\bibnamefont {Davinson}}, \bibinfo {author}
  {\bibfnamefont {P.~J.}\ \bibnamefont {Woods}}, \bibinfo {author}
  {\bibfnamefont {A.}~\bibnamefont {Jokinen}}, \bibinfo {author} {\bibfnamefont
  {H.}~\bibnamefont {Penttila}}, \bibinfo {author} {\bibfnamefont
  {F.}~\bibnamefont {Haas}}, \ and\ \bibinfo {author} {\bibfnamefont
  {S.}~\bibnamefont {Courtin}},\ }\href {\doibase 10.1103/PhysRevC.86.064308}
  {\bibfield  {journal} {\bibinfo  {journal} {Phys. Rev. C}\ }\textbf {\bibinfo
  {volume} {86}},\ \bibinfo {pages} {064308} (\bibinfo {year}
  {2012})}\BibitemShut {NoStop}%
\bibitem [{\citenamefont {Taniguchi}\ \emph {et~al.}(2009)\citenamefont
  {Taniguchi}, \citenamefont {Kanada-En'yo},\ and\ \citenamefont
  {Kimura}}]{PhysRevC.80.044316}%
  \BibitemOpen
  \bibfield  {author} {\bibinfo {author} {\bibfnamefont {Y.}~\bibnamefont
  {Taniguchi}}, \bibinfo {author} {\bibfnamefont {Y.}~\bibnamefont
  {Kanada-En'yo}}, \ and\ \bibinfo {author} {\bibfnamefont {M.}~\bibnamefont
  {Kimura}},\ }\href {\doibase 10.1103/PhysRevC.80.044316} {\bibfield
  {journal} {\bibinfo  {journal} {Phys. Rev. C}\ }\textbf {\bibinfo {volume}
  {80}},\ \bibinfo {pages} {044316} (\bibinfo {year} {2009})}\BibitemShut
  {NoStop}%
\bibitem [{\citenamefont {Darai}\ \emph {et~al.}(2012)\citenamefont {Darai},
  \citenamefont {Cseh},\ and\ \citenamefont {Jenkins}}]{PhysRevC.86.064309}%
  \BibitemOpen
  \bibfield  {author} {\bibinfo {author} {\bibfnamefont {J.}~\bibnamefont
  {Darai}}, \bibinfo {author} {\bibfnamefont {J.}~\bibnamefont {Cseh}}, \ and\
  \bibinfo {author} {\bibfnamefont {D.~G.}\ \bibnamefont {Jenkins}},\ }\href
  {\doibase 10.1103/PhysRevC.86.064309} {\bibfield  {journal} {\bibinfo
  {journal} {Phys. Rev. C}\ }\textbf {\bibinfo {volume} {86}},\ \bibinfo
  {pages} {064309} (\bibinfo {year} {2012})}\BibitemShut {NoStop}%
\bibitem [{Kim()}]{Kimura_private_comm}%
  \BibitemOpen
  \href@noop {} {}\bibinfo {note} {M. Kimura, Private
  Communication}\BibitemShut {NoStop}%
\bibitem [{\citenamefont {Cseh}\ and\ \citenamefont
  {Riczu}(2016)}]{Cseh2016312}%
  \BibitemOpen
  \bibfield  {author} {\bibinfo {author} {\bibfnamefont {J.}~\bibnamefont
  {Cseh}}\ and\ \bibinfo {author} {\bibfnamefont {G.}~\bibnamefont {Riczu}},\
  }\href {\doibase http://dx.doi.org/10.1016/j.physletb.2016.03.080} {\bibfield
   {journal} {\bibinfo  {journal} {Physics Letters B}\ }\textbf {\bibinfo
  {volume} {757}},\ \bibinfo {pages} {312 } (\bibinfo {year}
  {2016})}\BibitemShut {NoStop}%
\bibitem [{\citenamefont {Brenneisen}\ \emph
  {et~al.}(1995{\natexlab{a}})\citenamefont {Brenneisen}, \citenamefont
  {Grathwohl}, \citenamefont {Lickert}, \citenamefont {Ott}, \citenamefont
  {R{\"o}pke}, \citenamefont {Schm{\"a}lzlin}, \citenamefont {Siedle},\ and\
  \citenamefont {Wildenthal}}]{BrenneisenIII}%
  \BibitemOpen
  \bibfield  {author} {\bibinfo {author} {\bibfnamefont {J.}~\bibnamefont
  {Brenneisen}}, \bibinfo {author} {\bibfnamefont {D.}~\bibnamefont
  {Grathwohl}}, \bibinfo {author} {\bibfnamefont {M.}~\bibnamefont {Lickert}},
  \bibinfo {author} {\bibfnamefont {R.}~\bibnamefont {Ott}}, \bibinfo {author}
  {\bibfnamefont {H.}~\bibnamefont {R{\"o}pke}}, \bibinfo {author}
  {\bibfnamefont {J.}~\bibnamefont {Schm{\"a}lzlin}}, \bibinfo {author}
  {\bibfnamefont {P.}~\bibnamefont {Siedle}}, \ and\ \bibinfo {author}
  {\bibfnamefont {B.~H.}\ \bibnamefont {Wildenthal}},\ }\href {\doibase
  10.1007/BF01299758} {\bibfield  {journal} {\bibinfo  {journal} {Zeitschrift
  f{\"u}r Physik A Hadrons and Nuclei}\ }\textbf {\bibinfo {volume} {352}},\
  \bibinfo {pages} {403} (\bibinfo {year} {1995}{\natexlab{a}})}\BibitemShut
  {NoStop}%
\bibitem [{\citenamefont {Cseh}(1994)}]{PhysRevC.50.2240}%
  \BibitemOpen
  \bibfield  {author} {\bibinfo {author} {\bibfnamefont {J.}~\bibnamefont
  {Cseh}},\ }\href {\doibase 10.1103/PhysRevC.50.2240} {\bibfield  {journal}
  {\bibinfo  {journal} {Phys. Rev. C}\ }\textbf {\bibinfo {volume} {50}},\
  \bibinfo {pages} {2240} (\bibinfo {year} {1994})}\BibitemShut {NoStop}%
\bibitem [{\citenamefont {Neveling}\ \emph {et~al.}(2011)\citenamefont
  {Neveling}, \citenamefont {Fujita}, \citenamefont {Smit}, \citenamefont
  {Adachi}, \citenamefont {Berg}, \citenamefont {Buthelezi}, \citenamefont
  {Carter}, \citenamefont {Conradie}, \citenamefont {Couder}, \citenamefont
  {Fearick}, \citenamefont {Förtsch}, \citenamefont {Fourie}, \citenamefont
  {Fujita}, \citenamefont {Görres}, \citenamefont {Hatanaka}, \citenamefont
  {Jingo}, \citenamefont {Krumbholz}, \citenamefont {Kureba}, \citenamefont
  {Mira}, \citenamefont {Murray}, \citenamefont {von Neumann-Cosel},
  \citenamefont {O'Brien}, \citenamefont {Papka}, \citenamefont {Poltoratska},
  \citenamefont {Richter}, \citenamefont {Sideras-Haddad}, \citenamefont
  {Swartz}, \citenamefont {Tamii}, \citenamefont {Usman},\ and\ \citenamefont
  {van Zyl}}]{Neveling201129}%
  \BibitemOpen
  \bibfield  {author} {\bibinfo {author} {\bibfnamefont {R.}~\bibnamefont
  {Neveling}}, \bibinfo {author} {\bibfnamefont {H.}~\bibnamefont {Fujita}},
  \bibinfo {author} {\bibfnamefont {F.}~\bibnamefont {Smit}}, \bibinfo {author}
  {\bibfnamefont {T.}~\bibnamefont {Adachi}}, \bibinfo {author} {\bibfnamefont
  {G.}~\bibnamefont {Berg}}, \bibinfo {author} {\bibfnamefont {E.}~\bibnamefont
  {Buthelezi}}, \bibinfo {author} {\bibfnamefont {J.}~\bibnamefont {Carter}},
  \bibinfo {author} {\bibfnamefont {J.}~\bibnamefont {Conradie}}, \bibinfo
  {author} {\bibfnamefont {M.}~\bibnamefont {Couder}}, \bibinfo {author}
  {\bibfnamefont {R.}~\bibnamefont {Fearick}}, \bibinfo {author} {\bibfnamefont
  {S.}~\bibnamefont {Förtsch}}, \bibinfo {author} {\bibfnamefont
  {D.}~\bibnamefont {Fourie}}, \bibinfo {author} {\bibfnamefont
  {Y.}~\bibnamefont {Fujita}}, \bibinfo {author} {\bibfnamefont
  {J.}~\bibnamefont {Görres}}, \bibinfo {author} {\bibfnamefont
  {K.}~\bibnamefont {Hatanaka}}, \bibinfo {author} {\bibfnamefont
  {M.}~\bibnamefont {Jingo}}, \bibinfo {author} {\bibfnamefont
  {A.}~\bibnamefont {Krumbholz}}, \bibinfo {author} {\bibfnamefont
  {C.}~\bibnamefont {Kureba}}, \bibinfo {author} {\bibfnamefont
  {J.}~\bibnamefont {Mira}}, \bibinfo {author} {\bibfnamefont {S.}~\bibnamefont
  {Murray}}, \bibinfo {author} {\bibfnamefont {P.}~\bibnamefont {von
  Neumann-Cosel}}, \bibinfo {author} {\bibfnamefont {S.}~\bibnamefont
  {O'Brien}}, \bibinfo {author} {\bibfnamefont {P.}~\bibnamefont {Papka}},
  \bibinfo {author} {\bibfnamefont {I.}~\bibnamefont {Poltoratska}}, \bibinfo
  {author} {\bibfnamefont {A.}~\bibnamefont {Richter}}, \bibinfo {author}
  {\bibfnamefont {E.}~\bibnamefont {Sideras-Haddad}}, \bibinfo {author}
  {\bibfnamefont {J.}~\bibnamefont {Swartz}}, \bibinfo {author} {\bibfnamefont
  {A.}~\bibnamefont {Tamii}}, \bibinfo {author} {\bibfnamefont
  {I.}~\bibnamefont {Usman}}, \ and\ \bibinfo {author} {\bibfnamefont
  {J.}~\bibnamefont {van Zyl}},\ }\href {\doibase
  http://dx.doi.org/10.1016/j.nima.2011.06.077} {\bibfield  {journal} {\bibinfo
   {journal} {{Nuclear Instruments and Methods in Physics Research Section A:
  Accelerators, Spectrometers, Detectors and Associated Equipment}}\ }\textbf
  {\bibinfo {volume} {654}},\ \bibinfo {pages} {29 } (\bibinfo {year}
  {2011})}\BibitemShut {NoStop}%
\bibitem [{\citenamefont {Tamii}\ \emph {et~al.}(2009)\citenamefont {Tamii},
  \citenamefont {Fujita}, \citenamefont {Matsubara}, \citenamefont {Adachi},
  \citenamefont {Carter}, \citenamefont {Dozono}, \citenamefont {Fujita},
  \citenamefont {Fujita}, \citenamefont {Hashimoto}, \citenamefont {Hatanaka},
  \citenamefont {Itahashi}, \citenamefont {Itoh}, \citenamefont {Kawabata},
  \citenamefont {Nakanishi}, \citenamefont {Ninomiya}, \citenamefont
  {Perez-Cerdan}, \citenamefont {Popescu}, \citenamefont {Rubio}, \citenamefont
  {Saito}, \citenamefont {Sakaguchi}, \citenamefont {Sakemi}, \citenamefont
  {Sasamoto}, \citenamefont {Shimbara}, \citenamefont {Shimizu}, \citenamefont
  {Smit}, \citenamefont {Tameshige}, \citenamefont {Yosoi},\ and\ \citenamefont
  {Zenhiro}}]{Tamii2009326}%
  \BibitemOpen
  \bibfield  {author} {\bibinfo {author} {\bibfnamefont {A.}~\bibnamefont
  {Tamii}}, \bibinfo {author} {\bibfnamefont {Y.}~\bibnamefont {Fujita}},
  \bibinfo {author} {\bibfnamefont {H.}~\bibnamefont {Matsubara}}, \bibinfo
  {author} {\bibfnamefont {T.}~\bibnamefont {Adachi}}, \bibinfo {author}
  {\bibfnamefont {J.}~\bibnamefont {Carter}}, \bibinfo {author} {\bibfnamefont
  {M.}~\bibnamefont {Dozono}}, \bibinfo {author} {\bibfnamefont
  {H.}~\bibnamefont {Fujita}}, \bibinfo {author} {\bibfnamefont
  {K.}~\bibnamefont {Fujita}}, \bibinfo {author} {\bibfnamefont
  {H.}~\bibnamefont {Hashimoto}}, \bibinfo {author} {\bibfnamefont
  {K.}~\bibnamefont {Hatanaka}}, \bibinfo {author} {\bibfnamefont
  {T.}~\bibnamefont {Itahashi}}, \bibinfo {author} {\bibfnamefont
  {M.}~\bibnamefont {Itoh}}, \bibinfo {author} {\bibfnamefont {T.}~\bibnamefont
  {Kawabata}}, \bibinfo {author} {\bibfnamefont {K.}~\bibnamefont {Nakanishi}},
  \bibinfo {author} {\bibfnamefont {S.}~\bibnamefont {Ninomiya}}, \bibinfo
  {author} {\bibfnamefont {A.}~\bibnamefont {Perez-Cerdan}}, \bibinfo {author}
  {\bibfnamefont {L.}~\bibnamefont {Popescu}}, \bibinfo {author} {\bibfnamefont
  {B.}~\bibnamefont {Rubio}}, \bibinfo {author} {\bibfnamefont
  {T.}~\bibnamefont {Saito}}, \bibinfo {author} {\bibfnamefont
  {H.}~\bibnamefont {Sakaguchi}}, \bibinfo {author} {\bibfnamefont
  {Y.}~\bibnamefont {Sakemi}}, \bibinfo {author} {\bibfnamefont
  {Y.}~\bibnamefont {Sasamoto}}, \bibinfo {author} {\bibfnamefont
  {Y.}~\bibnamefont {Shimbara}}, \bibinfo {author} {\bibfnamefont
  {Y.}~\bibnamefont {Shimizu}}, \bibinfo {author} {\bibfnamefont
  {F.}~\bibnamefont {Smit}}, \bibinfo {author} {\bibfnamefont {Y.}~\bibnamefont
  {Tameshige}}, \bibinfo {author} {\bibfnamefont {M.}~\bibnamefont {Yosoi}}, \
  and\ \bibinfo {author} {\bibfnamefont {J.}~\bibnamefont {Zenhiro}},\ }\href
  {\doibase http://dx.doi.org/10.1016/j.nima.2009.03.248} {\bibfield  {journal}
  {\bibinfo  {journal} {Nuclear Instruments and Methods in Physics Research
  Section A: Accelerators, Spectrometers, Detectors and Associated Equipment}\
  }\textbf {\bibinfo {volume} {605}},\ \bibinfo {pages} {326 } (\bibinfo {year}
  {2009})}\BibitemShut {NoStop}%
\bibitem [{\citenamefont {Fujita}(2004)}]{Hiro_Internal_Memo}%
  \BibitemOpen
  \bibfield  {author} {\bibinfo {author} {\bibfnamefont {H.}~\bibnamefont
  {Fujita}},\ }\href@noop {} {\enquote {\bibinfo {title} {K600 internal
  memo},}\ } (\bibinfo {year} {2004})\BibitemShut {NoStop}%
\bibitem [{\citenamefont {Lukyanov}\ \emph {et~al.}(2006)\citenamefont
  {Lukyanov}, \citenamefont {Zemlyanaya},\ and\ \citenamefont
  {Lukyanov}}]{Lukyanov2006}%
  \BibitemOpen
  \bibfield  {author} {\bibinfo {author} {\bibfnamefont {V.~K.}\ \bibnamefont
  {Lukyanov}}, \bibinfo {author} {\bibfnamefont {E.~V.}\ \bibnamefont
  {Zemlyanaya}}, \ and\ \bibinfo {author} {\bibfnamefont {K.~V.}\ \bibnamefont
  {Lukyanov}},\ }\href {\doibase 10.1134/S1063778806020086} {\bibfield
  {journal} {\bibinfo  {journal} {Physics of Atomic Nuclei}\ }\textbf {\bibinfo
  {volume} {69}},\ \bibinfo {pages} {240} (\bibinfo {year} {2006})}\BibitemShut
  {NoStop}%
\bibitem [{ENS(2015)}]{ENSDF}%
  \BibitemOpen
  \href@noop {} {\enquote {\bibinfo {title} {{ENSDF, NNDC online data service,
  ENSDF database}},}\ }\bibinfo {howpublished}
  {{\url{http://www.nndc.bnl.gov/ensdf/}}} (\bibinfo {year} {2015})\BibitemShut
  {NoStop}%
\bibitem [{\citenamefont {Endt}(1990)}]{Endt19901}%
  \BibitemOpen
  \bibfield  {author} {\bibinfo {author} {\bibfnamefont {P.}~\bibnamefont
  {Endt}},\ }\href {\doibase http://dx.doi.org/10.1016/0375-9474(90)90598-G}
  {\bibfield  {journal} {\bibinfo  {journal} {Nuclear Physics A}\ }\textbf
  {\bibinfo {volume} {521}},\ \bibinfo {pages} {1 } (\bibinfo {year}
  {1990})}\BibitemShut {NoStop}%
\bibitem [{\citenamefont {{Li}}\ \emph {et~al.}(2016)\citenamefont {{Li}},
  \citenamefont {{Neveling}}, \citenamefont {{Adsley}}, \citenamefont
  {{Papka}}, \citenamefont {{Smit}}, \citenamefont {{Br{\"u}mmer}},
  \citenamefont {{Diget}}, \citenamefont {{Freer}}, \citenamefont {{Harakeh}},
  \citenamefont {{Kokalova}}, \citenamefont {{Nemulodi}}, \citenamefont
  {{Pellegri}}, \citenamefont {{Rebeiro}}, \citenamefont {{Swartz}},
  \citenamefont {{Triambak}}, \citenamefont {{van Zyl}},\ and\ \citenamefont
  {{Wheldon}}}]{2016arXiv161007437L}%
  \BibitemOpen
  \bibfield  {author} {\bibinfo {author} {\bibfnamefont {K.~C.~W.}\
  \bibnamefont {{Li}}}, \bibinfo {author} {\bibfnamefont {R.}~\bibnamefont
  {{Neveling}}}, \bibinfo {author} {\bibfnamefont {P.}~\bibnamefont
  {{Adsley}}}, \bibinfo {author} {\bibfnamefont {P.}~\bibnamefont {{Papka}}},
  \bibinfo {author} {\bibfnamefont {F.~D.}\ \bibnamefont {{Smit}}}, \bibinfo
  {author} {\bibfnamefont {J.~W.}\ \bibnamefont {{Br{\"u}mmer}}}, \bibinfo
  {author} {\bibfnamefont {C.~A.}\ \bibnamefont {{Diget}}}, \bibinfo {author}
  {\bibfnamefont {M.}~\bibnamefont {{Freer}}}, \bibinfo {author} {\bibfnamefont
  {M.~N.}\ \bibnamefont {{Harakeh}}}, \bibinfo {author} {\bibfnamefont
  {T.}~\bibnamefont {{Kokalova}}}, \bibinfo {author} {\bibfnamefont
  {F.}~\bibnamefont {{Nemulodi}}}, \bibinfo {author} {\bibfnamefont
  {L.}~\bibnamefont {{Pellegri}}}, \bibinfo {author} {\bibfnamefont
  {B.}~\bibnamefont {{Rebeiro}}}, \bibinfo {author} {\bibfnamefont {J.~A.}\
  \bibnamefont {{Swartz}}}, \bibinfo {author} {\bibfnamefont {S.}~\bibnamefont
  {{Triambak}}}, \bibinfo {author} {\bibfnamefont {J.~J.}\ \bibnamefont {{van
  Zyl}}}, \ and\ \bibinfo {author} {\bibfnamefont {C.}~\bibnamefont
  {{Wheldon}}},\ }\href@noop {} {\bibfield  {journal} {\bibinfo  {journal}
  {ArXiv e-prints}\ } (\bibinfo {year} {2016})},\ \Eprint
  {http://arxiv.org/abs/1610.07437} {arXiv:1610.07437 [nucl-ex]} \BibitemShut
  {NoStop}%
\bibitem [{\citenamefont {{Li}}(2015)}]{KCWLThesis}%
  \BibitemOpen
  \bibfield  {author} {\bibinfo {author} {\bibfnamefont {K.~C.~W.}\
  \bibnamefont {{Li}}},\ }\emph {\bibinfo {title} {{Characterization of the
  pre-eminent 4-$\alpha$ cluster state candiate in $^{16}$O}}},\ \href
  {http://hdl.handle.net/10019.1/97982} {Master's thesis},\ \bibinfo  {school}
  {Stellenbosch University} (\bibinfo {year} {2015})\BibitemShut {NoStop}%
\bibitem [{\citenamefont {Schneider}\ \emph {et~al.}(1979)\citenamefont
  {Schneider}, \citenamefont {Richter}, \citenamefont {Schwierczinski},
  \citenamefont {Spamer}, \citenamefont {Titze},\ and\ \citenamefont
  {Knüpfer}}]{Schneider197913}%
  \BibitemOpen
  \bibfield  {author} {\bibinfo {author} {\bibfnamefont {R.}~\bibnamefont
  {Schneider}}, \bibinfo {author} {\bibfnamefont {A.}~\bibnamefont {Richter}},
  \bibinfo {author} {\bibfnamefont {A.}~\bibnamefont {Schwierczinski}},
  \bibinfo {author} {\bibfnamefont {E.}~\bibnamefont {Spamer}}, \bibinfo
  {author} {\bibfnamefont {O.}~\bibnamefont {Titze}}, \ and\ \bibinfo {author}
  {\bibfnamefont {W.}~\bibnamefont {Knüpfer}},\ }\href {\doibase
  http://dx.doi.org/10.1016/0375-9474(79)90413-5} {\bibfield  {journal}
  {\bibinfo  {journal} {Nuclear Physics A}\ }\textbf {\bibinfo {volume}
  {323}},\ \bibinfo {pages} {13 } (\bibinfo {year} {1979})}\BibitemShut
  {NoStop}%
\bibitem [{\citenamefont {Brenneisen}\ \emph
  {et~al.}(1995{\natexlab{b}})\citenamefont {Brenneisen}, \citenamefont
  {Grathwohl}, \citenamefont {Lickert}, \citenamefont {Ott}, \citenamefont
  {Röpke}, \citenamefont {Schmälzlin}, \citenamefont {Siedle},\ and\
  \citenamefont {Wildenthal}}]{BrenneisenI}%
  \BibitemOpen
  \bibfield  {author} {\bibinfo {author} {\bibfnamefont {J.}~\bibnamefont
  {Brenneisen}}, \bibinfo {author} {\bibfnamefont {D.}~\bibnamefont
  {Grathwohl}}, \bibinfo {author} {\bibfnamefont {M.}~\bibnamefont {Lickert}},
  \bibinfo {author} {\bibfnamefont {R.}~\bibnamefont {Ott}}, \bibinfo {author}
  {\bibfnamefont {H.}~\bibnamefont {Röpke}}, \bibinfo {author} {\bibfnamefont
  {J.}~\bibnamefont {Schmälzlin}}, \bibinfo {author} {\bibfnamefont
  {P.}~\bibnamefont {Siedle}}, \ and\ \bibinfo {author} {\bibfnamefont
  {B.}~\bibnamefont {Wildenthal}},\ }\href {\doibase 10.1007/BF01298901}
  {\bibfield  {journal} {\bibinfo  {journal} {Zeitschrift für Physik A Hadrons
  and Nuclei}\ }\textbf {\bibinfo {volume} {352}},\ \bibinfo {pages} {149}
  (\bibinfo {year} {1995}{\natexlab{b}})}\BibitemShut {NoStop}%
\bibitem [{\citenamefont {Brenneisen}\ \emph
  {et~al.}(1995{\natexlab{c}})\citenamefont {Brenneisen}, \citenamefont
  {Grathwohl}, \citenamefont {Lickert}, \citenamefont {Ott}, \citenamefont
  {Röpke}, \citenamefont {Schmälzlin}, \citenamefont {Siedle},\ and\
  \citenamefont {Wildenthal}}]{BrenneisenII}%
  \BibitemOpen
  \bibfield  {author} {\bibinfo {author} {\bibfnamefont {J.}~\bibnamefont
  {Brenneisen}}, \bibinfo {author} {\bibfnamefont {D.}~\bibnamefont
  {Grathwohl}}, \bibinfo {author} {\bibfnamefont {M.}~\bibnamefont {Lickert}},
  \bibinfo {author} {\bibfnamefont {R.}~\bibnamefont {Ott}}, \bibinfo {author}
  {\bibfnamefont {H.}~\bibnamefont {Röpke}}, \bibinfo {author} {\bibfnamefont
  {J.}~\bibnamefont {Schmälzlin}}, \bibinfo {author} {\bibfnamefont
  {P.}~\bibnamefont {Siedle}}, \ and\ \bibinfo {author} {\bibfnamefont
  {B.}~\bibnamefont {Wildenthal}},\ }\href {\doibase 10.1007/BF01289501}
  {\bibfield  {journal} {\bibinfo  {journal} {Zeitschrift für Physik A Hadrons
  and Nuclei}\ }\textbf {\bibinfo {volume} {352}},\ \bibinfo {pages} {279}
  (\bibinfo {year} {1995}{\natexlab{c}})}\BibitemShut {NoStop}%
\bibitem [{\citenamefont {Strandberg}\ \emph {et~al.}(2008)\citenamefont
  {Strandberg}, \citenamefont {Beard}, \citenamefont {Couder}, \citenamefont
  {Couture}, \citenamefont {Falahat}, \citenamefont {G\"orres}, \citenamefont
  {LeBlanc}, \citenamefont {Lee}, \citenamefont {O'Brien}, \citenamefont
  {Palumbo}, \citenamefont {Stech}, \citenamefont {Tan}, \citenamefont
  {Ugalde}, \citenamefont {Wiescher}, \citenamefont {Costantini}, \citenamefont
  {Scheller}, \citenamefont {Pignatari}, \citenamefont {Azuma},\ and\
  \citenamefont {Buchmann}}]{PhysRevC.77.055801}%
  \BibitemOpen
  \bibfield  {author} {\bibinfo {author} {\bibfnamefont {E.}~\bibnamefont
  {Strandberg}}, \bibinfo {author} {\bibfnamefont {M.}~\bibnamefont {Beard}},
  \bibinfo {author} {\bibfnamefont {M.}~\bibnamefont {Couder}}, \bibinfo
  {author} {\bibfnamefont {A.}~\bibnamefont {Couture}}, \bibinfo {author}
  {\bibfnamefont {S.}~\bibnamefont {Falahat}}, \bibinfo {author} {\bibfnamefont
  {J.}~\bibnamefont {G\"orres}}, \bibinfo {author} {\bibfnamefont {P.~J.}\
  \bibnamefont {LeBlanc}}, \bibinfo {author} {\bibfnamefont {H.~Y.}\
  \bibnamefont {Lee}}, \bibinfo {author} {\bibfnamefont {S.}~\bibnamefont
  {O'Brien}}, \bibinfo {author} {\bibfnamefont {A.}~\bibnamefont {Palumbo}},
  \bibinfo {author} {\bibfnamefont {E.}~\bibnamefont {Stech}}, \bibinfo
  {author} {\bibfnamefont {W.~P.}\ \bibnamefont {Tan}}, \bibinfo {author}
  {\bibfnamefont {C.}~\bibnamefont {Ugalde}}, \bibinfo {author} {\bibfnamefont
  {M.}~\bibnamefont {Wiescher}}, \bibinfo {author} {\bibfnamefont
  {H.}~\bibnamefont {Costantini}}, \bibinfo {author} {\bibfnamefont
  {K.}~\bibnamefont {Scheller}}, \bibinfo {author} {\bibfnamefont
  {M.}~\bibnamefont {Pignatari}}, \bibinfo {author} {\bibfnamefont
  {R.}~\bibnamefont {Azuma}}, \ and\ \bibinfo {author} {\bibfnamefont
  {L.}~\bibnamefont {Buchmann}},\ }\href {\doibase 10.1103/PhysRevC.77.055801}
  {\bibfield  {journal} {\bibinfo  {journal} {Phys. Rev. C}\ }\textbf {\bibinfo
  {volume} {77}},\ \bibinfo {pages} {055801} (\bibinfo {year}
  {2008})}\BibitemShut {NoStop}%
\bibitem [{\citenamefont {Kawabata}\ \emph {et~al.}(2007)\citenamefont
  {Kawabata}, \citenamefont {Akimune}, \citenamefont {Fujita}, \citenamefont
  {Fujita}, \citenamefont {Fujiwara}, \citenamefont {Hara}, \citenamefont
  {Hatanaka}, \citenamefont {Itoh}, \citenamefont {Kanada-En'yo}, \citenamefont
  {Kishi}, \citenamefont {Nakanishi}, \citenamefont {Sakaguchi}, \citenamefont
  {Shimbara}, \citenamefont {Tamii}, \citenamefont {Terashima}, \citenamefont
  {Uchida}, \citenamefont {Wakasa}, \citenamefont {Yasuda}, \citenamefont
  {Yoshida},\ and\ \citenamefont {Yosoi}}]{Kawabata20076}%
  \BibitemOpen
  \bibfield  {author} {\bibinfo {author} {\bibfnamefont {T.}~\bibnamefont
  {Kawabata}}, \bibinfo {author} {\bibfnamefont {H.}~\bibnamefont {Akimune}},
  \bibinfo {author} {\bibfnamefont {H.}~\bibnamefont {Fujita}}, \bibinfo
  {author} {\bibfnamefont {Y.}~\bibnamefont {Fujita}}, \bibinfo {author}
  {\bibfnamefont {M.}~\bibnamefont {Fujiwara}}, \bibinfo {author}
  {\bibfnamefont {K.}~\bibnamefont {Hara}}, \bibinfo {author} {\bibfnamefont
  {K.}~\bibnamefont {Hatanaka}}, \bibinfo {author} {\bibfnamefont
  {M.}~\bibnamefont {Itoh}}, \bibinfo {author} {\bibfnamefont {Y.}~\bibnamefont
  {Kanada-En'yo}}, \bibinfo {author} {\bibfnamefont {S.}~\bibnamefont {Kishi}},
  \bibinfo {author} {\bibfnamefont {K.}~\bibnamefont {Nakanishi}}, \bibinfo
  {author} {\bibfnamefont {H.}~\bibnamefont {Sakaguchi}}, \bibinfo {author}
  {\bibfnamefont {Y.}~\bibnamefont {Shimbara}}, \bibinfo {author}
  {\bibfnamefont {A.}~\bibnamefont {Tamii}}, \bibinfo {author} {\bibfnamefont
  {S.}~\bibnamefont {Terashima}}, \bibinfo {author} {\bibfnamefont
  {M.}~\bibnamefont {Uchida}}, \bibinfo {author} {\bibfnamefont
  {T.}~\bibnamefont {Wakasa}}, \bibinfo {author} {\bibfnamefont
  {Y.}~\bibnamefont {Yasuda}}, \bibinfo {author} {\bibfnamefont
  {H.}~\bibnamefont {Yoshida}}, \ and\ \bibinfo {author} {\bibfnamefont
  {M.}~\bibnamefont {Yosoi}},\ }\href {\doibase
  http://dx.doi.org/10.1016/j.physletb.2006.11.079} {\bibfield  {journal}
  {\bibinfo  {journal} {Physics Letters B}\ }\textbf {\bibinfo {volume}
  {646}},\ \bibinfo {pages} {6 } (\bibinfo {year} {2007})}\BibitemShut
  {NoStop}%
\bibitem [{\citenamefont {Cseh}\ \emph {et~al.}(1982)\citenamefont {Cseh},
  \citenamefont {Koltay}, \citenamefont {Máté}, \citenamefont {Somorjai},\
  and\ \citenamefont {Zolnai}}]{Cseh198243}%
  \BibitemOpen
  \bibfield  {author} {\bibinfo {author} {\bibfnamefont {J.}~\bibnamefont
  {Cseh}}, \bibinfo {author} {\bibfnamefont {E.}~\bibnamefont {Koltay}},
  \bibinfo {author} {\bibfnamefont {Z.}~\bibnamefont {Máté}}, \bibinfo
  {author} {\bibfnamefont {E.}~\bibnamefont {Somorjai}}, \ and\ \bibinfo
  {author} {\bibfnamefont {L.}~\bibnamefont {Zolnai}},\ }\href {\doibase
  http://dx.doi.org/10.1016/0375-9474(82)90488-2} {\bibfield  {journal}
  {\bibinfo  {journal} {Nuclear Physics A}\ }\textbf {\bibinfo {volume}
  {385}},\ \bibinfo {pages} {43 } (\bibinfo {year} {1982})}\BibitemShut
  {NoStop}%
\bibitem [{\citenamefont {Cseh}\ and\ \citenamefont {Kat\ifmmode~\bar{o}\else
  \={o}\fi{}}(2013)}]{PhysRevC.87.067301}%
  \BibitemOpen
  \bibfield  {author} {\bibinfo {author} {\bibfnamefont {J.}~\bibnamefont
  {Cseh}}\ and\ \bibinfo {author} {\bibfnamefont {K.}~\bibnamefont
  {Kat\ifmmode~\bar{o}\else \={o}\fi{}}},\ }\href {\doibase
  10.1103/PhysRevC.87.067301} {\bibfield  {journal} {\bibinfo  {journal} {Phys.
  Rev. C}\ }\textbf {\bibinfo {volume} {87}},\ \bibinfo {pages} {067301}
  (\bibinfo {year} {2013})}\BibitemShut {NoStop}%
\bibitem [{\citenamefont {Cseh}(2015)}]{Cseh2015213}%
  \BibitemOpen
  \bibfield  {author} {\bibinfo {author} {\bibfnamefont {J.}~\bibnamefont
  {Cseh}},\ }\href {\doibase http://dx.doi.org/10.1016/j.physletb.2015.02.034}
  {\bibfield  {journal} {\bibinfo  {journal} {Physics Letters B}\ }\textbf
  {\bibinfo {volume} {743}},\ \bibinfo {pages} {213 } (\bibinfo {year}
  {2015})}\BibitemShut {NoStop}%
\bibitem [{\citenamefont {Dytrych}\ \emph {et~al.}(2008)\citenamefont
  {Dytrych}, \citenamefont {Sviratcheva}, \citenamefont {Draayer},
  \citenamefont {Bahri},\ and\ \citenamefont {Vary}}]{Dytrych2008}%
  \BibitemOpen
  \bibfield  {author} {\bibinfo {author} {\bibfnamefont {T.}~\bibnamefont
  {Dytrych}}, \bibinfo {author} {\bibfnamefont {K.~D.}\ \bibnamefont
  {Sviratcheva}}, \bibinfo {author} {\bibfnamefont {J.~P.}\ \bibnamefont
  {Draayer}}, \bibinfo {author} {\bibfnamefont {C.}~\bibnamefont {Bahri}}, \
  and\ \bibinfo {author} {\bibfnamefont {J.~P.}\ \bibnamefont {Vary}},\ }\href
  {http://stacks.iop.org/0954-3899/35/i=12/a=123101} {\bibfield  {journal}
  {\bibinfo  {journal} {Journal of Physics G: Nuclear and Particle Physics}\
  }\textbf {\bibinfo {volume} {35}},\ \bibinfo {pages} {123101} (\bibinfo
  {year} {2008})}\BibitemShut {NoStop}%
\bibitem [{\citenamefont {Arima}\ \emph {et~al.}(1970)\citenamefont {Arima},
  \citenamefont {Gillet},\ and\ \citenamefont
  {Ginocchio}}]{PhysRevLett.25.1043}%
  \BibitemOpen
  \bibfield  {author} {\bibinfo {author} {\bibfnamefont {A.}~\bibnamefont
  {Arima}}, \bibinfo {author} {\bibfnamefont {V.}~\bibnamefont {Gillet}}, \
  and\ \bibinfo {author} {\bibfnamefont {J.}~\bibnamefont {Ginocchio}},\ }\href
  {\doibase 10.1103/PhysRevLett.25.1043} {\bibfield  {journal} {\bibinfo
  {journal} {Phys. Rev. Lett.}\ }\textbf {\bibinfo {volume} {25}},\ \bibinfo
  {pages} {1043} (\bibinfo {year} {1970})}\BibitemShut {NoStop}%
\bibitem [{\citenamefont {Wigner}(1937)}]{PhysRev.51.106}%
  \BibitemOpen
  \bibfield  {author} {\bibinfo {author} {\bibfnamefont {E.}~\bibnamefont
  {Wigner}},\ }\href {\doibase 10.1103/PhysRev.51.106} {\bibfield  {journal}
  {\bibinfo  {journal} {Phys. Rev.}\ }\textbf {\bibinfo {volume} {51}},\
  \bibinfo {pages} {106} (\bibinfo {year} {1937})}\BibitemShut {NoStop}%
\bibitem [{\citenamefont {Elliott}(1958)}]{Elliott128}%
  \BibitemOpen
  \bibfield  {author} {\bibinfo {author} {\bibfnamefont {J.~P.}\ \bibnamefont
  {Elliott}},\ }\href {\doibase 10.1098/rspa.1958.0072} {\bibfield  {journal}
  {\bibinfo  {journal} {Proceedings of the Royal Society of London A:
  Mathematical, Physical and Engineering Sciences}\ }\textbf {\bibinfo {volume}
  {245}},\ \bibinfo {pages} {128} (\bibinfo {year} {1958})}\BibitemShut
  {NoStop}%
\bibitem [{\citenamefont {Harvey}(1973)}]{HARVEY1973191}%
  \BibitemOpen
  \bibfield  {author} {\bibinfo {author} {\bibfnamefont {M.}~\bibnamefont
  {Harvey}},\ }\href {\doibase http://dx.doi.org/10.1016/0375-9474(73)90251-0}
  {\bibfield  {journal} {\bibinfo  {journal} {Nuclear Physics A}\ }\textbf
  {\bibinfo {volume} {202}},\ \bibinfo {pages} {191 } (\bibinfo {year}
  {1973})}\BibitemShut {NoStop}%
\bibitem [{\citenamefont {Cseh}(1992)}]{CSEH1992173}%
  \BibitemOpen
  \bibfield  {author} {\bibinfo {author} {\bibfnamefont {J.}~\bibnamefont
  {Cseh}},\ }\href {\doibase http://dx.doi.org/10.1016/0370-2693(92)91124-R}
  {\bibfield  {journal} {\bibinfo  {journal} {Physics Letters B}\ }\textbf
  {\bibinfo {volume} {281}},\ \bibinfo {pages} {173 } (\bibinfo {year}
  {1992})}\BibitemShut {NoStop}%
\bibitem [{\citenamefont {Cseh}\ and\ \citenamefont
  {Levai}(1994)}]{CSEH1994165}%
  \BibitemOpen
  \bibfield  {author} {\bibinfo {author} {\bibfnamefont {J.}~\bibnamefont
  {Cseh}}\ and\ \bibinfo {author} {\bibfnamefont {G.}~\bibnamefont {Levai}},\
  }\href {\doibase http://dx.doi.org/10.1006/aphy.1994.1024} {\bibfield
  {journal} {\bibinfo  {journal} {Annals of Physics}\ }\textbf {\bibinfo
  {volume} {230}},\ \bibinfo {pages} {165 } (\bibinfo {year}
  {1994})}\BibitemShut {NoStop}%
\bibitem [{\citenamefont {Iachello}\ and\ \citenamefont
  {Levine}(1982)}]{IachelloJChemPhys_77_3046}%
  \BibitemOpen
  \bibfield  {author} {\bibinfo {author} {\bibfnamefont {F.}~\bibnamefont
  {Iachello}}\ and\ \bibinfo {author} {\bibfnamefont {R.~D.}\ \bibnamefont
  {Levine}},\ }\href@noop {} {\bibfield  {journal} {\bibinfo  {journal} {The
  Journal of Chemical Physics}\ }\textbf {\bibinfo {volume} {77}} (\bibinfo
  {year} {1982})}\BibitemShut {NoStop}%
\bibitem [{\citenamefont {Sheline}\ \emph {et~al.}(1982)\citenamefont
  {Sheline}, \citenamefont {Kubono}, \citenamefont {Morita},\ and\
  \citenamefont {Tanaka}}]{Sheline1982263}%
  \BibitemOpen
  \bibfield  {author} {\bibinfo {author} {\bibfnamefont {R.~K.}\ \bibnamefont
  {Sheline}}, \bibinfo {author} {\bibfnamefont {S.}~\bibnamefont {Kubono}},
  \bibinfo {author} {\bibfnamefont {K.}~\bibnamefont {Morita}}, \ and\ \bibinfo
  {author} {\bibfnamefont {M.}~\bibnamefont {Tanaka}},\ }\href {\doibase
  http://dx.doi.org/10.1016/0370-2693(82)90666-9} {\bibfield  {journal}
  {\bibinfo  {journal} {Physics Letters B}\ }\textbf {\bibinfo {volume}
  {119}},\ \bibinfo {pages} {263 } (\bibinfo {year} {1982})}\BibitemShut
  {NoStop}%
\bibitem [{\citenamefont {Blomqvist}\ and\ \citenamefont
  {Molinari}(1968)}]{BLOMQVIST1968545}%
  \BibitemOpen
  \bibfield  {author} {\bibinfo {author} {\bibfnamefont {J.}~\bibnamefont
  {Blomqvist}}\ and\ \bibinfo {author} {\bibfnamefont {A.}~\bibnamefont
  {Molinari}},\ }\href {\doibase
  http://dx.doi.org/10.1016/0375-9474(68)90515-0} {\bibfield  {journal}
  {\bibinfo  {journal} {Nuclear Physics A}\ }\textbf {\bibinfo {volume}
  {106}},\ \bibinfo {pages} {545 } (\bibinfo {year} {1968})}\BibitemShut
  {NoStop}%
\bibitem [{\citenamefont {Avila}\ \emph {et~al.}(2014)\citenamefont {Avila},
  \citenamefont {Rogachev}, \citenamefont {Goldberg}, \citenamefont {Johnson},
  \citenamefont {Kemper}, \citenamefont {Tchuvil'sky},\ and\ \citenamefont
  {Volya}}]{PhysRevC.90.024327}%
  \BibitemOpen
  \bibfield  {author} {\bibinfo {author} {\bibfnamefont {M.~L.}\ \bibnamefont
  {Avila}}, \bibinfo {author} {\bibfnamefont {G.~V.}\ \bibnamefont {Rogachev}},
  \bibinfo {author} {\bibfnamefont {V.~Z.}\ \bibnamefont {Goldberg}}, \bibinfo
  {author} {\bibfnamefont {E.~D.}\ \bibnamefont {Johnson}}, \bibinfo {author}
  {\bibfnamefont {K.~W.}\ \bibnamefont {Kemper}}, \bibinfo {author}
  {\bibfnamefont {Y.~M.}\ \bibnamefont {Tchuvil'sky}}, \ and\ \bibinfo {author}
  {\bibfnamefont {A.~S.}\ \bibnamefont {Volya}},\ }\href {\doibase
  10.1103/PhysRevC.90.024327} {\bibfield  {journal} {\bibinfo  {journal} {Phys.
  Rev. C}\ }\textbf {\bibinfo {volume} {90}},\ \bibinfo {pages} {024327}
  (\bibinfo {year} {2014})}\BibitemShut {NoStop}%
\bibitem [{\citenamefont {Wiedenh\"over}\ \emph {et~al.}(2001)\citenamefont
  {Wiedenh\"over}, \citenamefont {Wuosmaa}, \citenamefont {Janssens},
  \citenamefont {Lister}, \citenamefont {Carpenter}, \citenamefont {Amro},
  \citenamefont {Bhattacharyya}, \citenamefont {Brown}, \citenamefont
  {Caggiano}, \citenamefont {Devlin}, \citenamefont {Heinz}, \citenamefont
  {Kondev}, \citenamefont {Lauritsen}, \citenamefont {Sarantites},
  \citenamefont {Siem}, \citenamefont {Sobotka},\ and\ \citenamefont
  {Sonzogni}}]{PhysRevLett.87.142502}%
  \BibitemOpen
  \bibfield  {author} {\bibinfo {author} {\bibfnamefont {I.}~\bibnamefont
  {Wiedenh\"over}}, \bibinfo {author} {\bibfnamefont {A.~H.}\ \bibnamefont
  {Wuosmaa}}, \bibinfo {author} {\bibfnamefont {R.~V.~F.}\ \bibnamefont
  {Janssens}}, \bibinfo {author} {\bibfnamefont {C.~J.}\ \bibnamefont
  {Lister}}, \bibinfo {author} {\bibfnamefont {M.~P.}\ \bibnamefont
  {Carpenter}}, \bibinfo {author} {\bibfnamefont {H.}~\bibnamefont {Amro}},
  \bibinfo {author} {\bibfnamefont {P.}~\bibnamefont {Bhattacharyya}}, \bibinfo
  {author} {\bibfnamefont {B.~A.}\ \bibnamefont {Brown}}, \bibinfo {author}
  {\bibfnamefont {J.}~\bibnamefont {Caggiano}}, \bibinfo {author}
  {\bibfnamefont {M.}~\bibnamefont {Devlin}}, \bibinfo {author} {\bibfnamefont
  {A.}~\bibnamefont {Heinz}}, \bibinfo {author} {\bibfnamefont {F.~G.}\
  \bibnamefont {Kondev}}, \bibinfo {author} {\bibfnamefont {T.}~\bibnamefont
  {Lauritsen}}, \bibinfo {author} {\bibfnamefont {D.~G.}\ \bibnamefont
  {Sarantites}}, \bibinfo {author} {\bibfnamefont {S.}~\bibnamefont {Siem}},
  \bibinfo {author} {\bibfnamefont {L.~G.}\ \bibnamefont {Sobotka}}, \ and\
  \bibinfo {author} {\bibfnamefont {A.}~\bibnamefont {Sonzogni}},\ }\href
  {\doibase 10.1103/PhysRevLett.87.142502} {\bibfield  {journal} {\bibinfo
  {journal} {Phys. Rev. Lett.}\ }\textbf {\bibinfo {volume} {87}},\ \bibinfo
  {pages} {142502} (\bibinfo {year} {2001})}\BibitemShut {NoStop}%
\end{thebibliography}%

\end{document}